\documentclass[graybox]{svmult}

\smartqed
\usepackage{mathptmx}       
\usepackage{helvet}         
\usepackage{courier}        
\usepackage{type1cm}        
\usepackage{makeidx}         
\usepackage{graphicx}        
\usepackage{multicol}        
\usepackage[bottom]{footmisc}

\usepackage{amsmath}

\spnewtheorem{assumption}{Assumption}{\bf}{\it}

\makeindex             

\usepackage{amsfonts,amssymb}
\usepackage{graphicx,float,latexsym,subeqnarray,colordvi,epstopdf}

\def\AA{{\bf A}}
\def\BB{{\bf B}}
\def\CC{{\bf C}}
\def\DD{{\bf D}}
\def\FF{{\bf F}}
\def\II{{\bf I}}
\def\JJ{{\bf J}}
\def\LL{{\bf L}}
\def\PP{{\bf P}}
\def\UU{{\bf U}}

\def\R{\mathbb{R}}
\newcommand{\iin}{\rm{int}}
\newcommand{\lleft}{\rm{left}}
\newcommand{\rright}{\rm{right}}

\usepackage{color}

\begin{document}

\title*{Application of Operator Splitting Methods in Finance}
\author{Karel in 't Hout \and Jari Toivanen}
\institute{Karel in 't Hout \at Department of Mathematics and
Computer Science, University of Antwerp, Middelheimlaan 1,
B-2020 Antwerp, Belgium, \email{karel.inthout@uantwerp.be}
\and Jari Toivanen \at Institute for Computational and Mathematical
Engineering, Stanford University, Stanford, CA 94305, USA, \\
Department of Mathematical Information Technology, 
FI-40014 University of Jyv\"askyl\"a, Finland,
\email{toivanen@stanford.edu, jari.toivanen@jyu.fi}}
%
%
\maketitle

\abstract{
Financial derivatives pricing aims to find the fair value of a
financial contract on an underlying asset. Here we consider
option pricing in the partial differential equations framework. 
The contemporary models lead to one-dimensional or multidimensional parabolic 
problems of the convection-diffusion type and generalizations thereof.
An overview of various operator splitting methods is presented for the
efficient numerical solution of these problems.
\newline\indent
Splitting schemes of the Alternating Direction Implicit (ADI) type are
discussed for multidimensional problems, e.g.~given by stochastic volatility
(SV) models. For jump models Implicit-Explicit (IMEX) methods are considered
which efficiently treat the nonlocal jump operator.
For American options an easy-to-implement operator splitting method is
described for the resulting linear complementarity problems.
\newline\indent
Numerical experiments are presented to illustrate the actual stability and
convergence of the splitting schemes.
Here European and American put options are considered under four asset price 
models: the classical Black--Scholes model, the Merton jump-diffusion model,
the Heston SV model, and the Bates SV model with jumps.
}

\newpage
\section{Introduction}
In the contemporary international financial markets option products are widely traded. The average daily turnover in the global over-the-counter derivatives markets is huge. For example, in the foreign exchange market this was approximately equal to 337 billion US dollars in April 2013 \cite{BIS13}. In addition to standard call and put options, the so-called vanilla options, a broad range of exotic derivatives exists. One of the primary goals of financial mathematics is to determine the fair values of these derivatives as well as their sensitivities to underlying variables and parameters, which are crucial for hedging. To this purpose, advanced mathematical models are employed nowadays, yielding initial-boundary value problems for time-dependent partial differential equations (PDEs) and generalizations thereof, see 
e.g.~\cite{Andersen10,Clark11,Lipton01,Shreve08,Tavella00,Wilmott98}. These problems are in general multidimensional and of the convection-diffusion kind. In some cases analytical formulas in semi-closed form for the exact solutions have been obtained in the literature. For the majority of option valuation problems, however, such formulas are not available. In view of this, one resorts to numerical methods for their approximate solution. To banks and other financial institutions, the efficient, stable, and robust numerical approximation of option values and their sensitivities is of paramount importance.

A well-known and versatile approach to the numerical solution of time-dependent convection-diffusion equations is given by the {\it method of lines}. It consists of two general, consecutive steps. In the first step the PDE is discretized in the spatial variables, e.g. by finite difference, finite volume, or finite element methods. This leads to a so-called semidiscrete system of ordinary differential equations. In the second step the obtained semidiscrete system is numerically solved by applying a suitable, implicit time-discretization method. If the PDE is multidimensional, then the latter task can be computationally very intensive when standard application of classical implicit methods, such as the Crank--Nicolson scheme, is used. In the recent years, a variety of operator splitting methods have been developed that enable a highly efficient and stable numerical solution of semidiscretized multidimensional PDEs and generalizations thereof that arise in financial mathematics.

The aim of this chapter to give an overview of main classes of operator splitting methods with applications in finance. Here we have chosen to consider a variety of, increasingly sophisticated, models that are well-known in the financial option valuation literature.

We deal in the following with two basic types of options, involving a given so-called strike price $K>0$ and a given maturity time $T>0$, where today is always denoted by time~$0$.
A {\it European call (put) option} is a contract between two parties, the holder and the writer, which gives the holder the right to buy from (sell to) the writer a prescribed asset for the price $K$ at the future date $T$.
An {\it American call (put) option} is the same, except that the holder can exercise at any time between today and the maturity date.
An option is a right and not an obligation.
The underlying asset can be a stock, a foreign currency, a commodity, etc.
For a detailed introduction to financial options we refer to \cite{Hull11}.
Clearly, an option has value and a central question in
financial mathematics is what its fair value is.

\section{Models for Underlying Assets}\label{models}
\subsection{Geometric Brownian Motion}
The seminal papers by Black \& Scholes \cite{Black73} and Merton \cite{Merton73} 
present a key equation for the fair values of European call and put options.   
In these papers the dynamics of the underlying asset price is modeled by the 
stochastic differential equation (SDE)
\begin{equation}\label{GeomBrownSDE}
dS(t) = \mu S(t) dt + \sigma S(t) dW(t) \quad (t\ge 0).
\end{equation}
Here $W(t)$ denotes the Wiener process or standard Brownian motion, and $\mu$, $\sigma$ 
are given real parameters that are called the drift and the volatility, respectively.
The volatility is a degree for the uncertainty of the return realized on the asset. 

The SDE \eqref{GeomBrownSDE} describes a so-called geometric Brownian 
motion, which satisfies $S(t)\ge 0$ whenever $S(0)\ge 0$. 
Under this asset price model and several additional assumptions, Black,
Scholes, and Merton derived the famous partial
differential equation (PDE)
\begin{equation}\label{BSPDE}
\frac{\partial u}{\partial t}
= \frac{1}{2} \sigma^2 s^2 \frac{\partial^2 u}{\partial s^2}
+ r s \frac{\partial u}{\partial s} - r u \quad (s>0,\,\, 0<t \le T).
\end{equation}
Here $u(s,t)$ represents the fair value at time $T-t$ of a European vanilla
option if $S(T-t)=s$. 
The quantity $r$ in \eqref{BSPDE} is the risk-free interest rate and is given.
A main consequence of the Black, Scholes, and Merton analysis
is that the drift $\mu$ actually does not appear in the option pricing PDE.
This observation has led to the important risk-neutral valuation theory.
It is beyond the scope of the present chapter to discuss this theory in more
detail, but see e.g.~\cite{Hull11,Shreve08}.

In formulating \eqref{BSPDE} we have chosen $t$ as the time till maturity.
Thus the time runs in the opposite direction compared to \eqref{GeomBrownSDE}.
Accordingly, the payoff function $\phi$, which defines the value of the option 
contract at maturity time $T$, leads to an {\it initial condition}
\begin{equation}\label{BSIC}
u(s,0) = \phi (s) \quad (s\ge 0).
\end{equation}
For a European vanilla option with given strike price $K$ there holds
\begin{equation}\label{payoff}
\phi (s) =
\left\{
  \begin{array}{ll}
    \max (s-K, 0)\quad {\rm for}~~s\ge 0 & ~~\hbox{(call),} \\
    \max (K-s, 0)\quad {\rm for}~~s\ge 0 & ~~\hbox{(put),}
  \end{array}
\right.
\end{equation}
and at $s=0$ one has the Dirichlet boundary condition
\begin{equation}\label{BSBC}
u(0,t) =
\left\{
  \begin{array}{ll}
    0~~~~~~~\,\quad {\rm for}~~0\le t\le T & ~~\hbox{(call),} \\
    e^{-rt}K \quad {\rm for}~~0\le t\le T & ~~\hbox{(put).}
  \end{array}
\right.
\end{equation} 

Equation \eqref{BSPDE} is called the {\it Black--Scholes PDE}
or {\it Black--Scholes--Merton PDE}.
It is fully deterministic and it can be viewed as a time-dependent 
convection-diffusion-reaction equation.
For European vanilla options, an analytical solution $u$ in semi-closed form 
was derived in \cite{Black73}, constituting the well-known Black--Scholes 
formula.

The Black--Scholes PDE is generic in the sense that it is valid for a wide
range of European-style options. 
The initial and boundary conditions are determined by the specific option.
As an example, for a European up-and-out call option with given barrier 
$B>K$, the PDE \eqref{BSPDE} holds whenever $0<s<B,\,\, 0<t \le T$.
In this case, the initial condition is 
\begin{equation*}
u(s,0)= \max (s-K,0)\quad {\rm for}~~0\le s<B
\end{equation*}
and one has the Dirichlet boundary conditions 
\begin{equation*}
u(0,t)=u(B,t)=0 \quad {\rm for}~~0\le t\le T.
\end{equation*}
The homogeneous condition at $s=B$ corresponds to the fact that, by construction, 
an up-and-out call option becomes worthless whenever the underlying asset price 
moves above the barrier.

For many types of options, including (continuous) barrier options, semi-analytical 
pricing formulas have been obtained in the literature in the Black--Scholes framework, 
see e.g. \cite{Hull11}.
At present it is well-known, however, that each of the assumptions underlying this
framework are violated to a smaller or larger extent in practice.
In particular, the interest rate $r$ and the volatility $\sigma$ are not constant, 
but vary in time. 
In view of this, more advanced asset pricing models have been developed and, as a 
consequence, more advanced option valuation PDEs are obtained.
In this chapter we do not enter into the details of the mathematical connection
between asset price SDEs and option valuation PDEs, but mention that a main
tool is the celebrated  Feynman--Kac theorem, see e.g. \cite{Shreve08}.
In the following we discuss typical, contemporary instances of more advanced 
option valuation PDEs.

\subsection{Stochastic Volatility and Stochastic Interest Rate Models}\label{H_PDEs}
Heston \cite{Heston93} modeled the volatility itself by a SDE. 
The Heston stochastic volatility model is popular especially in the foreign exchange markets.
The corresponding option valuation PDE is
\begin{equation}\label{HPDE}
\frac{\partial u}{\partial t} = \tfrac{1}{2} s^2 v \frac{\partial^2 u}{\partial s^2}
+ \rho \sigma s v \frac{\partial^2 u}{\partial s \partial v}
+ \tfrac{1}{2}\sigma^2 v \frac{\partial^2 u}{\partial v^2}
+ r s \frac{\partial u}{\partial s} +
\kappa ( \eta - v) \frac{\partial u}{\partial v} - r u
\end{equation}
for $s>0$, $v>0$, and $0<t\leq T$.
Here $u(s,v,t)$ represents the fair value of a European-style option if at $t$
time units before maturity the asset price equals $s$ and
the variance equals $v$.
We note that by definition the variance is the square of the volatility.
The positive parameters $\kappa$ and $\eta$ are the mean-reversion rate
and long-term mean, respectively, of the variance,
$\sigma>0$ is the volatility-of-variance, and $\rho \in [-1,1]$ denotes
the correlation between the two underlying Brownian motions.
Equation \eqref{HPDE} is called the {\it Heston PDE}. 
It can be viewed as a time-dependent convection-diffusion-reaction equation on an
unbounded, two-dimensional spatial domain.
If the correlation $\rho$ is nonzero, which almost always holds in practice, then the Heston PDE contains a mixed spatial derivative term.

For a European vanilla option under the Heston model, one has an initial condition as well as a boundary condition at $s=0$ that are the same as in the Black--Scholes case discussed above.
In the Heston case there is also a boundary $v=0$. 
Observe that as $v\downarrow 0$, then all second-order derivative terms vanish in \eqref{HPDE}.
It has been proved in \cite{Ekstrom10} that for the fair option value 
function $u$ the Heston PDE is fulfilled if $v=0$, which constitutes the (nonstandard) 
boundary condition at $v=0$.

For the Heston asset pricing model (which we did not explicitly formulate) the so-called 
Feller condition $2\kappa\eta \ge \sigma^2$ is often considered in the literature.
This condition determines whether or not the variance process can attain the value zero 
(given a strictly positive initial variance): it cannot attain zero if and only if 
Feller holds.
The situation where the Feller condition is violated is well-known to be challenging when 
numerically solving the Heston asset pricing model.
For the Heston option valuation PDE \eqref{HPDE}, on the other hand, it turns out that this 
issue is not critical in the numerical solution.

A refinement of the Heston model is obtained by considering also a stochastic interest rate,
see e.g. \cite{Grzelak11,Grzelak12,Haentjens13a,Haentjens12}.
As an illustration we consider the case where the interest rate is described 
by the well-known Hull--White model \cite{Hull11,Hull90}.
This leads to the following so-called {\it Heston--Hull--White (HHW) PDE} for the option 
value function $u = u(s,v,r,t)$:
\begin{align}\label{HHWPDE}
\frac{\partial u}{\partial t} =
&~\tfrac{1}{2}s^2v\frac{\partial^2 u}{\partial s^2}
+ \tfrac{1}{2}\sigma_1^2 v \frac{\partial^2 u}{\partial v^2}
+ \tfrac{1}{2}\sigma_2^2\frac{\partial^2 u}{\partial r^2}
+ \rho_{12} \sigma_1 s v \frac{\partial^2 u}{\partial s \partial v}
+ \rho_{13} \sigma_2 s \sqrt{v} \frac{\partial^2 u}{\partial s \partial r}\nonumber\\
&~
+ \rho_{23} \sigma_1 \sigma_2 \sqrt{v} \frac{\partial^2 u}{\partial v \partial r}
+ rs \frac{\partial u}{\partial s} + \kappa (\eta - v)\frac{\partial u}{\partial v}
+ a(b(T-t)-r)\frac{\partial u}{\partial r} - ru
\end{align}
for $s>0$, $v>0$, $-\infty < r < \infty$, and $0< t \le T$.
Here $\kappa$, $\eta$, $\sigma_1$, $a$, and $\sigma_2$ are given positive real
constants and $b$ denotes a given deterministic, positive function of time.
Further, there are given correlations $\rho_{12}$, $\rho_{13}$, $\rho_{23}\in [-1,1]$.
Clearly, the HHW PDE is a time-dependent convection-diffusion-reaction equation on an unbounded, three-dimensional spatial domain with three mixed derivative terms.
For a European vanilla option, initial and boundary conditions are the same as in the Heston case above.
Note that if $v\downarrow 0$, then all second-order derivative terms, apart from the $\partial^2 u/ \partial r^2$ term, vanish in \eqref{HHWPDE}.

The Heston and HHW models are two of many instances of asset pricing models that lead to multidimensional option valuation PDEs.
Multidimensional PDEs are also obtained when considering other types of options, e.g. options on a basket of assets. 
Then, in the Black--Scholes framework, the dimension of the PDE is equal to the number of assets.
In general, analytical solutions in (semi-)closed form to these PDEs are not available.

\subsection{Jump Models}\label{jumpmodels}
Sometimes the value of the underlying asset changes so rapidly that this
would have very tiny probability under the above Brownian motion 
based models. For example, the stock price during a market crash or after
a major news event can move very fast. 
Already in 1976, Merton proposed in
\cite{Merton76} to add a jump component in the model of the underlying
asset price. 
In his model, the jumps are log-normally distributed and their arrival times
follow a Poisson process. After a jump the value of the asset
is obtained by multiplying the value before the jump by a random variable
with the probability density function (PDF)
\begin{equation}\label{MertonPDF}
f (y) = \frac{1}{y\delta \sqrt{2\pi}}
\exp{\left(-\frac{(\log y-\gamma)^2}{2\delta^2}\right)}
\end{equation}
for $y>0$, where $\gamma$ is the mean of the normal distribution and
$\delta$ is its standard deviation.
Kou proposed in \cite{Kou02} a log-double-exponential distribution
defined by the PDF
\begin{equation}\label{logdouble}
f (y) = \left\{ \begin{aligned}
    &q\alpha_2y^{\alpha_2-1}, &&0<y<1,\\
    &p\alpha_1y^{-\alpha_1-1}, &&y\ge 1,
  \end{aligned}\right.\\
\end{equation}
where $p, q, \alpha_1 > 1$, and $\alpha_2$ are positive constants such
that $p + q = 1$. These models have finite jump activity which is
denoted by $\lambda$ here. There are also many popular infinite jump
activity models like the CGMY model \cite{Carr02}. 
In the following we shall consider only finite activity models.

The value $u(s,t)$ of a European option satisfies the partial
integro-differential equation (PIDE)
\begin{equation}\label{PIDE}
\frac{\partial u}{\partial t}
= \tfrac{1}{2} \sigma^2 s^2 \frac{\partial^2 u}{\partial s^2}
+ (r - \lambda\zeta) s \frac{\partial u}{\partial s}
- (r + \lambda) u + \lambda \int_0^\infty u(sy,t) f(y) dy
\end{equation}
for $s > 0$ and $0 < t \le T$, where $\zeta$ is the mean jump 
size given by
\begin{equation}\label{zetadef}
\zeta = \int_0^\infty ( y - 1 ) f(y) dy.
\end{equation}
For the Merton and Kou models the mean jumps are
$\zeta = e^{\gamma + \delta^2/2} - 1$ and
$\zeta = \frac{q \alpha_2}{\alpha_2 + 1} + \frac{p \alpha_1}{\alpha_1 - 1} - 1$,
respectively.

Bates proposed to combine the Heston stochastic volatility model
and the Merton jump model in \cite{Bates96}. Under this model the value 
$u(s,v,t)$ of a European option satisfies the PIDE
\begin{equation}\label{BatesPIDE}
\begin{split}
\frac{\partial u}{\partial t} = {}&{} \tfrac{1}{2} s^2 v \frac{\partial^2 u}{\partial s^2}
+ \rho \sigma s v \frac{\partial^2 u}{\partial s \partial v}
+ \tfrac{1}{2}\sigma^2 v \frac{\partial^2 u}{\partial v^2}
+ (r - \lambda\zeta) s \frac{\partial u}{\partial s} +
\kappa ( \eta - v) \frac{\partial u}{\partial v} \\
& {} - (r + \lambda) u + \lambda \int_0^\infty u(sy,v,t) f(y) dy
\end{split}
\end{equation}
for $s>0$, $v>0$, and $0<t\leq T$, where the PDF $f$ 
is given by \eqref{MertonPDF}. 
For an extensive discussion on jump models in finance see e.g.~\cite{Cont04}.

\section{Linear Complementarity Problem for American Options}\label{American}
Unlike European-style options, American-style options can be exercised at any
time up to the maturity date. Hence, the fair value of an American option is
always greater than or equal to the instantaneous payoff,
\begin{equation}\label{EarlyExConstr}
u \ge \phi.
\end{equation}
Due to this early exercise constraint, the P(I)DE does not hold everywhere anymore.
Instead, a linear complementarity problem (LCP) or partial \mbox{(integro-)}differential
complementarity problem is obtained in general for the fair value of an American option:
\begin{equation}\label{LCP}
\left\{ \begin{aligned}
& \displaystyle
\frac{\partial u}{\partial t} \ge {\mathcal A} u, \qquad u \ge \phi, \\[1mm]
& \displaystyle
\left(\frac{\partial u}{\partial t} - {\mathcal A} u \right) (u - \phi) = 0,
\end{aligned} \right.
\end{equation}
where ${\mathcal A}$ stands for the pertinent spatial differential operator.
For example, for the Black--Scholes model,
\begin{equation*}
{\mathcal A} u = \frac{1}{2} \sigma^2 s^2 \frac{\partial^2 u}{\partial s^2}
+ r s \frac{\partial u}{\partial s} - r u.
\end{equation*}
The above inequalities and equation hold pointwise. The equation in
\eqref{LCP} is the complementarity condition. It states that at each
point one of the two inequalities has to be an equality.
The paper \cite{Huang98} discusses the LCP formulation for
American-style options under various asset price models
and studies the structure and properties of the obtained fully discrete LCPs.

We note that the penalty approach is a popular alternative for LCPs.
Here a penalty term is added to the P(I)DE for a European option
with the aim to enforce the early exercise constraint \eqref{EarlyExConstr}.
The resulting problems are nonlinear and their efficient numerical solution
is considered in \cite{Forsyth02}, for example. For several other alternative
formulations and approximations for LCPs, we refer to \cite{Toivanen10b}.

\section{Spatial Discretization}\label{spacediscr}
In this chapter we employ finite difference (FD) discretizations for the 
spatial derivatives. 
An alternative approach would be to use finite element discretizations; 
see e.g. \cite{Achdou05,Seydel12}.
It is common practice to first truncate the infinite $s$-domain
$[0,\infty)$ to $[0,S_{\max}]$ with a sufficiently large, real $S_{\max}$. 
Typically one wishes $S_{\max}$ to be such that the error caused by
this truncation is a small fraction of the error due to the discretization
of the differential (and integral) operators. 
Similarly, with multidimensional models including the variance $v$ or
the interest rate $r$, their corresponding infinite
domains are truncated to sufficiently large bounded domains.
The truncation requires additional boundary conditions to be specified.
For an actual choice of these conditions for the models considered in Sections 
\ref{models}, \ref{American} we refer to Section \ref{experiments}.

Let the grid in the $s$-direction be defined by the $m_1 + 1$
grid points $0 = s_0 < s_1 < \cdots < s_{m_1} = S_{\max}$.
The corresponding grid sizes are denoted by 
$\Delta s_i = s_i - s_{i-1}$,\, $i = 1, 2, \ldots, m_1$.
For multidimensional models, we use tensor product grids. 
For example, in the case of a stochastic volatility model, if a grid for
the variance $v$ is given by $0 = v_0 < v_1 < \cdots < v_{m_2} = V_{\max}$, 
then $(m_1 + 1) \times (m_2 + 1)$ spatial grid points
are defined by $(s_i,v_j)$ with $i = 0, 1, \ldots, m_1$ and
$j = 0, 1, \ldots, m_2$.
In financial applications nonuniform grids are often preferable
over uniform grids. The use of suitable nonuniform grids will be
illustrated in Section \ref{experiments}.

For discretizing the first derivative $\frac{\partial u_i}{\partial s}$
and the second derivative $\frac{\partial^2 u_i}{\partial s^2}$ at $s = s_i$, 
we employ in this chapter the well-known central FD schemes
\begin{equation}\label{cfd1}
\frac{\partial u_i}{\partial s}
\approx
\frac{-\Delta s_{i+1}}{\Delta s_i (\Delta s_i + \Delta s_{i+1})} u_{i-1}
+ \frac{\Delta s_{i+1} - \Delta s_i}{\Delta s_i \Delta s_{i+1}} u_i
+ \frac{\Delta s_i}{(\Delta s_i + \Delta s_{i+1}) \Delta s_{i+1}} u_{i+1}
\end{equation}
and
\begin{equation}\label{cfd2}
\frac{\partial^2 u_i}{\partial s^2}
\approx
\frac{2}{\Delta s_i (\Delta s_i + \Delta s_{i+1})} u_{i-1}
- \frac{2}{\Delta s_i \Delta s_{i+1}} u_i
+ \frac{2}{(\Delta s_i + \Delta s_{i+1}) \Delta s_{i+1}} u_{i+1}.
\end{equation}
With multidimensional models the analogous schemes are used
for the other spatial directions, thus e.g. for $\frac{\partial u_j}{\partial v}$ 
and $\frac{\partial^2 u_j}{\partial v^2}$ at $v = v_j$. 
For the mixed derivative 
$\frac{\partial^2 u_{i,j}}{\partial s\partial v}$ at $(s,v) = (s_i,v_j)$
we consider the 9-point stencil obtained by successively applying the central 
FD schemes for the first derivative in the $s$- and $v$-directions.
With sufficiently smooth varying grid sizes, the above
central FDs give second-order accurate approximations for
the derivatives. 

We mention that in financial applications other FD schemes are
employed as well, such as upwind discretization for first derivative terms 
or alternative discretizations for mixed derivative terms.

With the jump models the integral term needs to be discretized at grid points 
$s_i$. 
First the integral is divided into two parts
\begin{equation*}
\int_0^\infty u(s_i y,t) f(y) dy
= \int_0^{S_{\max}/s_i} u(s_i y,t) f(y) dy +
\int_{S_{\max}/s_i}^\infty u(s_i y,t) f(y) dy,
\end{equation*}
which correspond to the values of $u$ in the computational
domain $[0,S_{\max}]$ and outside of it, respectively. 
The second part can be estimated using knowledge 
about $u$ in the far field $[S_{\max},\infty)$.
For example, for put options $u$ is usually assumed to be
close to zero for $s \ge S_{\max}$ and, thus, the second integral
is approximated by zero in this case.
The PDFs $f$ are smooth functions apart from the potential jump at
$y=1$ in the Kou model. Due to the smoothness of the integrand the
trapezoidal rule leads to second-order accuracy with respect to
the grid size. This gives the approximation
\begin{equation*}
\int_0^{S_{\max}/s_i} u(s_i y,t) f(y) dy \approx
\sum_{j=1}^{m_1} \frac{\Delta s_j}{2 s_i} \left(
u(s_{j-1},t) f(s_{j-1}/s_i) + u(s_j,t) f(s_j/s_i) \right).
\end{equation*}
For example, the papers \cite{Salmi11} and \cite{Toivanen08} describe more
accurate quadrature rules for the Merton and Kou jumps models, respectively.
The discretization of the integral term leads to a dense matrix.
The integral can be transformed into a convolution integral
and due to this FFT can be used to compute it more efficiently;
see \cite{Almendral05,Andersen00,dHalluin05,Tavella00}, for example. 
In the case of the Kou model, efficient recursion
formulas can be used \cite{Carr07,Toivanen08}.

\section{Time Discretization}\label{timediscr}
\subsection{The $\theta$-method}
For any P(I)DE from Section~\ref{models}, the spatial discretization outlined
in Section~\ref{spacediscr} leads to an initial value problem for a system 
of ordinary differential equations,
\begin{equation}\label{ODE}
\dot{U}(t) = \AA(t) U(t) + G(t) \quad (0\le t\le T), \quad U(0) = U_0.
\end{equation}
Here $\AA(t)$ for $0\le t\le T$ is a given square real matrix and $G(t)$ is
a given real vector that depends on the boundary conditions.
The entries of the solution vector $U(t)$ represent approximations to the exact
solution of the option valuation P(I)DE at the spatial grid points, ordered in
a convenient way.
The vector $U_0$ is given by direct evaluation of the option's payoff function 
at these grid points.

The semidiscrete system \eqref{ODE} is stiff in general and, hence, implicit 
time discretization methods are natural candidates for its numerical solution.
Let parameter $\theta \in (0,1]$ be given.
Let time step $\Delta t =T/N$ with integer $N\ge 1$ and temporal grid points 
$t_n = n\, \Delta t$ for integers $0\le n\le N$.
The {\it $\theta$-method}\, forms a well-known implicit time discretization method. 
It generates approximations $U_n$ to $U(t_n)$ successively for $n=1,2,\ldots,N$ by
\begin{equation}\label{thetamethod}
U_n = U_{n-1} + (1-\theta) \Delta t\,\AA(t_{n-1})U_{n-1}
+ \theta \Delta t\, \AA(t_{n})U_n + \Delta t\, G_{n-1+\theta},
\end{equation}
where $G_{n-1+\theta}$ denotes an approximation to $G(t)$ at $t = (n-1+\theta)\Delta t$.
This can also be written as
\begin{equation*}
(\II-\theta \Delta t \AA(t_{n}))U_n = (\II + (1-\theta) \Delta t\,\AA(t_{n-1}))U_{n-1}
+ \Delta t\, G_{n-1+\theta},
\end{equation*}
with $\II$ the identity matrix of the same size as $\AA (t)$.
For \mbox{$\theta = 1$} one obtains the first-order {\it backward Euler method}\, and 
for $\theta = \frac{1}{2}$ the second-order {\it Crank--Nicolson method}\, or\, 
{\it trapezoidal rule}.
For simplicity we consider in this chapter only constant time steps,
but most of the presented time discretization methods can directly be 
extended to variable time steps.

When applying the Crank--Nicolson method, it is common practice in finance to 
first perform a few backward Euler steps to start the time stepping. 
This is often called Rannacher smoothing \cite{Rannacher84}. 
It helps to damp high-frequency components in the numerical solution, due to the 
nonsmooth initial (payoff) function, which are usually not sufficiently damped by 
the Crank--Nicolson method itself.

Clearly, in order to compute the vector $U_n$ defined by \eqref{thetamethod}, one 
has to solve a linear system of equations with the matrix $\II-\theta \Delta t \AA(t_{n})$.
When the option valuation PDE is multidimensional, the size of this matrix is usually 
very large and it possesses a large bandwidth. For a PIDE, this matrix is dense.
In these situations, the solution of the linear system can be computationally demanding 
when standard methods, like LU decomposition, are applied.
Time discretization methods based on operator splitting can then form an
attractive alternative.
The key idea is to split the matrix $\AA(t)$ into several parts, each of
which is  numerically handled more easily than the complete matrix itself.

\subsection{Operator Splitting Methods Based on Direction}\label{directionsplitting}
For multidimensional PDEs, splitting schemes of the Alternating Direction
Implicit (ADI) type are often applied in financial practice.
To illustrate the idea, the two-dimensional Heston PDE and three-dimensional
HHW PDE, given in Section \ref{H_PDEs}, are considered.
For the Heston PDE the semidiscrete system \eqref{ODE} is autonomous;
we split
\begin{equation*}
\AA = \AA_0 + \AA_1 + \AA_2.
\end{equation*}
Next, for the HHW PDE,
\begin{equation*}
\AA(t) = \AA_0 + \AA_1 + \AA_2 + \AA_3(t).
\end{equation*}
Here $\AA_0$ is chosen as the part that represents all mixed derivative
terms. It is nonzero whenever (one of) the correlation factor(s) is nonzero.
The parts $\AA_1$, $\AA_2$, and $\AA_3(t)$ represent
all spatial derivatives in the $s$-, $v$-, and $r$-directions, respectively.
The latter three matrices have, possibly up to permutation, all a fixed small
bandwidth. The vector $G(t)$ in the semidiscrete system is splitted in
a similar way. For notational convenience, define functions $\FF_j$ by
\begin{equation*}\label{Fj}
    \FF_j(t,V) = \AA_j V + G_j ~~(j=0,1,2) ~~ {\rm and} ~~
    \FF_3(t,V) = \AA_3(t) V + G_3(t)
\end{equation*}
for $0\le t\le T$, $V\in \R^m$.
Set $\FF = \sum_{j=0}^k \FF_j$ with $k=2$ for Heston and $k=3$ for HHW.
We discuss in this section four contemporary ADI-type splitting schemes:
\newpage
{\it Douglas (Do) scheme}
\begin{equation}\label{Do}
\left\{\begin{array}{lll}
Y_0 = U_{n-1}+\Delta t\, \FF(t_{n-1},U_{n-1}), \\\\
Y_j = Y_{j-1}+\theta\Delta t\, (\FF_j(t_n,Y_j)-\FF_j(t_{n-1},U_{n-1}))
\quad (j=1,2,\ldots,k), \\\\
U_n = Y_k.
\end{array}\right.
\end{equation}
{\it Craig--Sneyd (CS) scheme}
\begin{equation}\label{CS}
\left\{\begin{array}{lll}
Y_0 = U_{n-1}+\Delta t\, \FF(t_{n-1},U_{n-1}), \\\\
Y_j = Y_{j-1}+\theta\Delta t\, (\FF_j(t_n,Y_j)-\FF_j(t_{n-1},U_{n-1}))
\quad (j=1,2,\ldots,k), \\\\
\widetilde{Y}_0 = Y_0+\frac{1}{2}\Delta t\, (\FF_0(t_n,Y_k)-\FF_0(t_{n-1},U_{n-1})),\\\\
\widetilde{Y}_j = \widetilde{Y}_{j-1}+\theta\Delta t\, (\FF_j(t_n,\widetilde{Y}_j)-\FF_j(t_{n-1},U_{n-1}))
\quad (j=1,2,\ldots,k), \\\\
U_n = \widetilde{Y}_k.
\end{array}\right.
\end{equation}
{\it Modified Craig--Sneyd (MCS) scheme}
\begin{equation}\label{MCS}
\left\{\begin{array}{lll}
Y_0 = U_{n-1}+\Delta t\, \FF(t_{n-1},U_{n-1}), \\\\
Y_j = Y_{j-1}+\theta\Delta t\, (\FF_j(t_n,Y_j)-\FF_j(t_{n-1},U_{n-1}))
\quad (j=1,2,\ldots,k), \\\\
\widehat{Y}_0 = Y_0+\theta\Delta t\, (\FF_0(t_n,Y_k)-\FF_0(t_{n-1},U_{n-1})),\\\\
\widetilde{Y}_0 = \widehat{Y}_0+(\frac{1}{2}-\theta)\Delta t\, (\FF(t_n,Y_k)-\FF(t_{n-1},U_{n-1})), \\\\
\widetilde{Y}_j = \widetilde{Y}_{j-1}+\theta\Delta t\, (\FF_j(t_n,\widetilde{Y}_j)-\FF_j(t_{n-1},U_{n-1}))
\quad (j=1,2,\ldots,k), \\\\
U_n = \widetilde{Y}_k.
\end{array}\right.
\end{equation}
{\it Hundsdorfer--Verwer (HV) scheme}
\begin{equation}\label{HV}
\left\{\begin{array}{lll}
Y_0 = U_{n-1}+\Delta t\, \FF(t_{n-1},U_{n-1}), \\\\
Y_j = Y_{j-1}+\theta\Delta t\, (\FF_j(t_n,Y_j)-\FF_j(t_{n-1},U_{n-1}))
\quad (j=1,2,\ldots,k), \\\\
\widetilde{Y}_0 = Y_0+\frac{1}{2}\Delta t\, (\FF(t_n,Y_k)-\FF(t_{n-1},U_{n-1})),\\\\
\widetilde{Y}_j = \widetilde{Y}_{j-1}+\theta\Delta t\, (\FF_j(t_n,\widetilde{Y}_j)-\FF_j(t_n,Y_k))
\quad (j=1,2,\ldots,k),\\\\
U_n = \widetilde{Y}_k.
\end{array}\right.
\end{equation}

In the Do scheme \eqref{Do}, a forward Euler predictor step is followed
by~$k$ implicit but unidirectional corrector steps that serve to stabilize
the predictor step. The CS scheme \eqref{CS}, the MCS scheme
\eqref{MCS}, and the HV scheme \eqref{HV} 
can be viewed as different extensions to the Do scheme.
Indeed, their first two lines are identical to those of the Do scheme.
They next all perform a second predictor step, followed 
by~$k$ unidirectional corrector steps.
Observe that the CS and MCS schemes are equivalent if (and only if) 
$\theta=\tfrac{1}{2}$.

Clearly, in all four ADI schemes the $\AA_0$ part, representing all mixed
derivatives, is always treated in an {\it explicit}\, fashion.
In the original formulation of ADI schemes mixed derivative terms were not 
considered.
It is a common and natural use in the literature to refer to the above, extended 
schemes also as ADI schemes.
In the special case where $\FF_0= 0$, the CS scheme reduces to the Do scheme,
but the MCS scheme (with $\theta \not= \frac{1}{2}$) and the HV scheme do not.
Following the original ADI approach, the $\AA_1$, $\AA_2$, $\AA_3(t)$ parts
are treated in an {\it implicit}\, fashion.
In every step of each scheme, systems of linear equations need to be
solved involving the matrices $(\II-\theta\, \Delta t\, \AA_j)$ for  
$j=1,2$ as well as $(\II-\theta\, \Delta t\, \AA_3(t_n))$ if $k=3$.
Since all these matrices have a fixed, small bandwidth, this can be
done very efficiently by means of LU decomposition, cf.~also
Section \ref{direct}.
Because for $j=1, 2$ the pertinent matrices are further independent 
of the step index $n$, their LU decompositions can be 
computed once, beforehand, and then used in all time steps.
Accordingly, for each ADI scheme, the number of floating point operations per 
time step is directly proportional to the number of spatial grid 
points, which is a highly favorable property.

By Taylor expansion one obtains (after some elaborate calculations) the
classical order of consistency\footnote{That is, the order for 
fixed nonstiff ODE systems.} of each ADI scheme.
For any given~$\theta$, the order of the Do scheme is just {\it one} 
whenever $\AA_0$ is nonzero.
This low order is due to the fact that the $\AA_0$ part is treated in a
simple, forward Euler fashion.
The CS scheme has order {\it two} provided $\theta=\tfrac{1}{2}$.
The MCS and HV schemes are of order two for any given $\theta$.
A virtue of ADI schemes, compared to other operator splitting schemes
based on direction, is that the internal vectors $Y_j$, 
$\widetilde{Y}_j$ form consistent approximations to $U(t_n)$.

The Do scheme can be regarded as a generalization of the original ADI schemes 
for two-dimensional diffusion equations by Douglas \& Rachford \cite{Douglas56} 
and Peaceman \& Rachford \cite{Peaceman55} to the situation where mixed 
derivative terms are present.
This generalization was first considered by McKee \& Mitchell \cite{McKee70} 
for diffusion equations and subsequently in \cite{McKee96}
for convection-diffusion equations.

The CS scheme was developed by Craig \& Sneyd \cite{Craig88} with the aim to 
obtain a stable second-order ADI scheme for diffusion equations with mixed 
derivative terms.

The MCS scheme was constructed by In 't Hout \& Welfert \cite{intHout09b}
so as to arrive at more freedom in the choice of $\theta$ as compared to 
the second-order CS scheme.

The HV scheme was designed by Hundsdorfer \cite{Hundsdorfer02} and Verwer et.~al.
\cite{Verwer99} for the numerical solution of convection-diffusion-reaction 
equations arising in atmospheric chemistry, cf.~also \cite{Hundsdorfer03}.
The application of the HV scheme to equations containing mixed derivative
terms was first studied in \cite{intHout07,intHout09b}.

The Do and CS schemes are well-known for PDEs in finance, see e.g.
\cite{Andersen10,Lipton01}. More recently, the MCS and HV schemes
have gained interest, see e.g.
\cite{Clark11,Dang10,Egloff11,Haentjens13a,Haentjens12,intHout10,Itkin11}.

The formulation of the ADI schemes \eqref{Do}--\eqref{HV} is analogous
to the type of formulation used in \cite{Hundsdorfer02}.
In the literature, ADI schemes are also sometimes referred to as Stabilizing 
Correction schemes, and are further closely related to Approximate Matrix 
Factorization methods and Implicit-Explicit (IMEX) Runge--Kutta methods, 
cf.~e.g.~\cite{Hundsdorfer03}.

In \cite{intHout11,intHout13,intHout07,intHout09b} comprehensive stability
results in the von Neumann sense have been derived for the four schemes
\eqref{Do}--\eqref{HV} in the application to multidimensional
convection-diffusion equations with mixed derivative terms.
These results concern unconditional stability, that is,
without any restriction on the time step $\Delta t$.
For each ADI scheme, lower bounds on $\theta$ guaranteeing unconditional
stability have been obtained, depending in particular on the spatial
dimension. Based on these theoretical stability results and the numerical
experience in \cite{Haentjens13a,Haentjens12,intHout10} 
the following values are found to be useful for $k=2, 3$:
\begin{itemize}
\item Do scheme~~~\, with $\theta = \frac{1}{2}$ (if $k=2$) and $\theta = \frac{2}{3}$ (if $k=3$)
\item CS scheme~~~\, with $\theta = \frac{1}{2}$
\item MCS scheme with $\theta = \frac{1}{3}$ (if $k=2$) and $\theta = \max\{\frac{1}{3},\frac{2}{13}(2\gamma+1)\}$ (if $k=3$)
\item HV scheme~~~ with $\theta = \frac{1}{2}+\frac{1}{6}\sqrt{3}$.
\end{itemize} 
Here $\gamma =\max \left\{ |\rho_{12}|, |\rho_{13}|, |\rho_{23}| \right\} \in [0,1]$,
which is a measure for the relative size of the mixed derivative coefficients.

In addition to ADI schemes, there exists a variety of well-known
alternative operator splitting schemes based on direction, called
Locally One-Dimensional (LOD) methods, fractional step methods,
or componentwise splitting schemes. These schemes originate in the
1960s in the work by Dyakonov, Marchuk, Samarskii, Yanenko,
and others. Some of them are related to Strang splitting schemes, developed
at the same time. For a general overview and analysis of such
methods we refer to \cite{Hundsdorfer03,Marchuk90}.
Applications in financial mathematics of these schemes are considered
in, for example, \cite{Ikonen07a,Toivanen10a}.

\subsection{Operator Splitting Methods Based on Operator Type}
For the jump models considered in Section \ref{jumpmodels} the semidiscrete 
matrix $\AA$ can be written in the form
\begin{equation}
\AA = \DD + \JJ,
\end{equation}
where $\DD$ and $\JJ$ correspond to the differential operator and integral
operator, respectively. The matrix $\DD$ is sparse while in general $\JJ$ is
a dense matrix or has dense blocks. In view of the different nature of these 
two matrices it can be preferable to employ an operator splitting method based 
on them. 

In \cite{Andersen00}, Andersen and Andreasen describe a generalized
$\theta$-method
\begin{equation}\label{genthetamethod}
\left( \II - \theta_D \Delta t \DD - \theta_J \Delta t \JJ \right) U_n
= \left( \II + \left(1 - \theta_D \right) \Delta t \DD
+ \left( 1 - \theta_J \right) \Delta t \JJ \right) U_{n-1}
\end{equation}
assuming here $G = 0$.
The standard choice $\theta_D = 1$ and $\theta_J = 0$ corresponds 
to the {\it IMEX Euler method}: it treats the stiff differential part
implicitly, using the backward Euler method, and the nonstiff integral part
explicitly, using the forward Euler method. This choice yields first-order
consistency. The benefit is that it is not necessary
to solve dense linear systems involving the matrix $\JJ$. Instead, in each
time step only one multiplication with $\JJ$ is required.
This approach has been considered and analysed in \cite{Cont05}.

In \cite{Feng08} an extrapolation approach is advocated based on the IMEX 
Euler method. Here approximations at a given fixed time are computed for
a decreasing sequence of step sizes and then linearly combined so as to
achieve a high order of accuracy. 

In \cite{Andersen00} second-order consistency is
obtained through an alternating treatment of the $\DD$ and $\JJ$ parts. 
They propose to take a $\Delta t/2$ substep with $\theta_D = 1$ and
$\theta_J = 0$ followed by a $\Delta t/2$ substep with $\theta_D = 0$ and
$\theta_J = 1$. Here linear systems involving the dense matrix $\JJ$ need
to be solved, for which the authors employ FFT.

In \cite{dHalluin05} the original $\theta$-method is analyzed, 
where the linear system in each time step is solved by applying a
fixed-point iteration on the jump part following an idea 
in \cite{Tavella00}.

The following, second-order {\it IMEX midpoint scheme} has 
been considered in e.g. \cite{Feng08,Kwon11a,Kwon11b,Salmi12},
\begin{equation}\label{IMEXmidpoint}
( \II - \Delta t \DD ) U_n =
( \II + \Delta t \DD ) U_{n-2}
+ 2 \Delta t \JJ U_{n-1}
+ 2 \Delta t G_{n-1}.
\end{equation} 
The scheme \eqref{IMEXmidpoint} can be viewed as obtained from the semidiscrete system 
\eqref{ODE} at $t_{n-1}$ by the approximations 
$\DD U_{n-1} \approx \tfrac{1}{2} \DD (U_n + U_{n-2})$ 
and $\dot{U}_{n-1} \approx \tfrac{1}{2\Delta t} (U_n - U_{n-2})$.
Two subsequent second-order IMEX methods are the 
{\it IMEX--CNAB scheme}
\begin{equation}\label{IMEXCNAB}
\left( \II - \tfrac{\Delta t}{2} \DD \right) U_n =
\left( \II + \tfrac{\Delta t}{2} \DD \right) U_{n-1}
+ \tfrac{\Delta t}{2} \JJ \left( 3 U_{n-1} - U_{n-2} \right)
+ \Delta t G_{n-1/2}
\end{equation}
and the {\it IMEX--BDF2 scheme}
\begin{equation}\label{IMEXBDF2}
\left( \tfrac{3}{2} \II - \Delta t \DD \right) U_n =
2 U_{n-1} - \tfrac{1}{2} U_{n-2}
+ \Delta t \JJ \left( 2 U_{n-1} - U_{n-2} \right)
+ \Delta t G_n.
\end{equation}
These schemes have recently been applied for option pricing in \cite{Salmi14}
and can be regarded as obtained by approximating the semidiscrete system
\eqref{ODE} at 
$t_{n-1/2} = \tfrac{1}{2}(t_n + t_{n-1})$ and at $t_n$, respectively.

The IMEX schemes \eqref{IMEXmidpoint}, \eqref{IMEXCNAB}, and
\eqref{IMEXBDF2} were studied in a general framework, without application to
option valuation, in \cite{Frank97}. Here it was noted that such schemes
can be considered as starting with an implicit method and then replacing
the nonstiff part of the implicit term by an explicit formula using
extrapolation based on previous time steps. An overview of IMEX methods
is given in \cite{Hundsdorfer03}.

In general, IMEX methods are only conditionally stable, that is,
they are stable for a sufficiently small time step $\Delta t$.
For example, the IMEX midpoint scheme \eqref{IMEXmidpoint} and
the IMEX--CNAB scheme \eqref{IMEXCNAB} are stable whenever
$\lambda \Delta t < 1$ and the $\lambda u$ term in \eqref{PIDE}
is included in $\DD$; see \cite{Salmi14}. Recall that $\lambda$
denotes the jump activity.

The schemes discussed in this section are of the linear multistep type. 
For IMEX schemes of Runge--Kutta type applied to jump models we mention
\cite{Briani07}.

\subsection{Operator Splitting Method for Linear Complementarity Problems}\label{LCPsplit}
The fully discrete LCPs obtained by spatial and temporal discretization
of \eqref{LCP} for American-style options are more difficult
to solve than the corresponding systems of linear equations for
the European-style counterparts. It is desirable to split these LCPs
into simpler subproblems. Here we describe the operator splitting
method considered in \cite{Ikonen04,Ikonen09} which was motivated
by splitting methods for incompressible flows
\cite{Chorin68,Glowinski86}. To this purpose, we reformulate
LCPs with Lagrange multipliers.

The $\theta$-method discretization \eqref{thetamethod} naturally
gives rise to the following, fully discrete LCP
\begin{equation}\label{DLCP}
\left\{ \begin{aligned}
& \displaystyle
\BB U_n - \CC U_{n-1} - \Delta t G_{n-1+\theta} \ge 0,\\[1mm]
& \displaystyle
U_n \ge U_0,\quad \left( \BB U_n - \CC U_{n-1} - \Delta t G_{n-1+\theta} \right)^T \left( U_n - U_0 \right) = 0,
\end{aligned} \right.
\end{equation}
where $\BB = \II - \theta \Delta t \AA$,
$\CC = \II + (1 - \theta) \Delta t \AA$, and
$\AA$ is assumed to be constant in time.
By introducing a Lagrange multiplier vector $\lambda_n$,
the LCP \eqref{DLCP} takes the equivalent form
\begin{equation}\label{LagrangeLCP}
\left\{ \begin{aligned}
& \displaystyle
\BB U_n - \CC U_{n-1} - \Delta t G_{n-1+\theta} = \Delta t \lambda_n \ge 0,\\[1mm] 
& \displaystyle
U_n \ge U_0,\quad \left( \lambda_n \right)^T \left( U_n - U_0 \right) = 0.
\end{aligned} \right.
\end{equation}
The basic idea of the operator splitting method proposed in \cite{Ikonen04}
is to decouple in \eqref{LagrangeLCP} the first line from the second line.
This is accomplished by approximating the Lagrange multiplier $\lambda_n$
in the first line by the previous Lagrange multiplier $\lambda_{n-1}$.
This leads to the system of linear equations
\begin{equation}\label{LCPsplit1}
\BB \widetilde{U}_n = \CC U_{n-1} + \Delta t G_{n-1+\theta}
+ \Delta t \lambda_{n-1}.
\end{equation}
After solving this system, the intermediate solution vector
$\widetilde{U}_n$ and the Lagrange multiplier $\lambda_n$ are updated
to satisfy the (spatially decoupled) equation and complementarity conditions
\begin{equation}\label{LCPsplit2}
\left\{ \begin{aligned}
& \displaystyle
U_n - \widetilde{U}_n = \Delta t (\lambda_n - \lambda_{n-1}), \\[1mm]
& \displaystyle
\lambda_n \ge 0, \quad U_n \ge U_0,\quad 
\displaystyle \left( \lambda_n \right)^T \left( U_n - U_0 \right) = 0.\\[1mm]
\end{aligned} \right.
\end{equation}
Thus, this operator splitting method for American options leads to
the solution of linear systems \eqref{LCPsplit1}, which are essentially
the same as for European options, and a simple update step \eqref{LCPsplit2}.
This update can be performed very fast, at each spatial grid point
independently, with the formula
\begin{equation}\label{update}
(U_{n,i}~,~\lambda_{n,i}) = \left\{ \begin{aligned}
& \displaystyle
\left( \widetilde{U}_{n,i} - \Delta t \lambda_{n-1,i}~,~0 \right), \qquad
\text{if }\; \widetilde{U}_{n,i} - \Delta t \lambda_{n-1,i} > U_{0,i}~, \\
& \displaystyle
\left( U_{0,i}~,~\lambda_{n-1,i} + \tfrac{1}{\Delta t}
\left( U_{0,i} - \widetilde{U}_{n,i} \right) \right),
\qquad \text{otherwise.}
\end{aligned} \right.
\end{equation}

The above operator splitting approach has been studied for more
advanced time discretization schemes of both linear multistep and
Runge--Kutta type in \cite{Ikonen04,Ikonen09}.
Moreover, it has recently been effectively combined with IMEX schemes
in \cite{Salmi12} for the case of jump models and with ADI schemes
in \cite{Haentjens15} for the case of the Heston model.
For instance, the pertinent adaptations of the IMEX--CNAB scheme and
the MCS scheme are
\begin{equation*}\label{IMEXCNAB_Amer}
\left( \II - \tfrac{\Delta t}{2} \DD \right) \widetilde{U}_n =
\left( \II + \tfrac{\Delta t}{2} \DD \right) U_{n-1}
+ \tfrac{\Delta t}{2} \JJ \left( 3 U_{n-1} - U_{n-2} \right)
+ \Delta t G_{n-1/2} + \Delta t\, \lambda_{n-1},
\end{equation*}
and
\begin{equation*}\label{MCS_Amer}
\left\{\begin{array}{lll}
Y_0 = U_{n-1}+\Delta t\, \FF(t_{n-1},U_{n-1}) + \Delta t\, \lambda_{n-1}, \\\\
Y_j = Y_{j-1}+\theta\Delta t\, (\FF_j(t_n,Y_j)-\FF_j(t_{n-1},U_{n-1}))
\quad (j=1,2,\ldots,k), \\\\
\widehat{Y}_0 = Y_0+\theta\Delta t\, (\FF_0(t_n,Y_k)-\FF_0(t_{n-1},U_{n-1})),\\\\
\widetilde{Y}_0 = \widehat{Y}_0+(\frac{1}{2}-\theta)\Delta t\, (\FF(t_n,Y_k)-\FF(t_{n-1},U_{n-1})), \\\\
\widetilde{Y}_j = \widetilde{Y}_{j-1}+\theta\Delta t\, (\FF_j(t_n,\widetilde{Y}_j)-\FF_j(t_{n-1},U_{n-1}))
\quad (j=1,2,\ldots,k), \\\\
\widetilde{U}_n = \widetilde{Y}_k,
\end{array}\right.
\end{equation*}
respectively,
followed by the update \eqref{update}. The other three ADI schemes from
Section \ref{directionsplitting} are adapted analogously. Note that 
only a $\Delta t \lambda_{n-1}$ term has been added to the first line
of the MCS scheme \eqref{MCS}. Accordingly,
like for the $\theta$-method, the amount of computational work per
time step is essentially the same as for the corresponding European-style
option.

\section{Solvers for Algebraic Systems}
The implicit time discretizations described in Section \ref{timediscr}
lead, in each time step, to systems of linear equations of the form
\begin{equation}\label{linsys}
\BB U = \Psi
\end{equation}
or LCPs of the form
\begin{equation}\label{LCPsys}
\left\{ \begin{aligned}
& \displaystyle
\BB U \ge \Psi,
\qquad U \ge \Phi, \\[1mm]
& \displaystyle
\left( \BB U - \Psi \right)^T \left( U - \Phi \right) = 0
\end{aligned} \right.
\end{equation}
with given matrix $\BB$ and given vectors $\Phi$, $\Psi$. For models without
jumps, semidiscretization by finite difference, finite volume, and
finite element methods yields sparse matrices $\BB$.
For one-dimensional models, the central FDs \eqref{cfd1}
and \eqref{cfd2} lead to tridiagonal $\BB$. For higher dimensional models
they give rise to matrices $\BB$ with a large bandwidth whenever classical
(non-splitted) time stepping schemes are applied. On the other hand, for
the operator splitting methods based on direction (cf. Section
\ref{directionsplitting}) one also acquires tridiagonal matrices
(possibly after renumbering the unknowns). Wider FD stencils
lead to additional nonzero diagonals. Time discretization of jump models
with an implicit treatment of jumps makes $\BB$ dense.

\subsection{Direct Methods}\label{direct}
The system of linear equations \eqref{linsys} can be solved by a direct
method using LU decomposition. This method first forms a lower triangular
matrix $\LL$ and an upper triangular matrix $\UU$ such that $\BB = \LL\UU$.
After this the solution vector $U$ is obtained by solving first 
$\LL V = \Psi$ and then $\UU U = V$. 

Let $m$ denote the dimension of the matrix $\BB$.
For tridiagonal $\BB$, or more generally matrices with a fixed
small bandwidth, the LU decomposition yields optimal 
computational cost in the sense that the number of floating point operations 
is of order $m$.
Hence, it is very efficient for one-dimensional models and for
higher-dimensional models when operator splitting schemes based on
direction are applied.

For two-dimensional models with classical time stepping schemes,
a LU decomposition can be formed by order $m^{3/2}$ 
floating point operations if a nested dissection method
can be used and then the computational cost of the solution is of
order $m \log m$, see \cite{Davis06,George73}.
For higher-dimensional models with classical time stepping schemes, the 
computational cost is less favorable.

For a general matrix $\BB$, solving the LCP \eqref{LCPsys} requires
iterative methods. However, in the special case that $\BB$ is
tridiagonal, the solution vector satisfies $U_i = \Phi_i$
($1 \le i \le i_0$),\, $U_i > \Phi_i$ ($i_0 < i \le m$) for
certain $i_0$ and some additional assumptions hold, the
{\it Brennan--Schwartz algorithm} \cite{Brennan77} gives a direct
method to solve the LCP; see also \cite{Achdou05,Ikonen07b,Jaillet90}.
After inverting the numbering of the unknowns to be from right to 
left, represented by a permutation matrix $\PP$, this algorithm
is equivalent to applying the LU decomposition method to the
corresponding linear system with matrix $\PP\BB\PP$ where the
projection step is carried out directly after computing 
each component in the back substitution step with $\UU$.
More precisely the back substitution step reads after the
renumbering of unknowns:
\begin{equation}
\left\{ \begin{aligned}
& \displaystyle
U_m = \max \{ V_m / \UU_{m,m}\,,\,\Phi_m\}, \\[1mm]
& \displaystyle
U_i = \max \{ (V_i - \UU_{i,i+1} U_{i+1}) / \UU_{i,i}\,,\, \Phi_i\}
\quad (i = m-1,m-2,\ldots,1).
\end{aligned} \right.
\end{equation}
The Brennan--Schwartz algorithm is essentially as fast as the LU decomposition
method for linear systems and, thus, it has optimal computational cost.

\subsection{Iterative Methods}
There are many iterative methods for solving systems of linear equations.
The two most important method categories are the stationary iterative
methods and the Krylov subspace methods. Well-known Krylov subspace
methods for the, typically unsymmetric, system \eqref{linsys} are
the generalized minimal residual (GMRES) method \cite{Saad86} and
the BiCGSTAB method \cite{VanderVorst92}. In the following we discuss
a stationary iterative method in some more detail which is familiar
in finance applications.
The {\it successive over-relaxation (SOR) method} reads
\begin{equation}\label{SOR}
U_i^{(k+1)} = U_i^{(k)}
+ \frac{\omega}{\BB_{i,i}} \left( \Psi_i
- \sum_{j=1}^{i-1} \BB_{i,j}^{} U_j^{(k+1)}
- \sum_{j=i}^m \BB_{i,j}^{} U_j^{(k)} \right)\quad
\end{equation}
for $i=1,2,\ldots,m,~ k=0,1,2,\ldots$, where $\omega$ is a relaxation parameter.
This method reduces to the Gauss--Seidel method in the case $\omega = 1$.
The convergence rate of the iteration \eqref{SOR} can be improved
significantly by an optimal choice of $\omega$. Still the number of
iterations to reach a given accuracy typically grows with $m$, that is,
when the spatial grid is refined the convergence slows down.

The SOR iteration can be generalized for LCPs by performing a projection
after each update \cite{Cryer71}; see also \cite{Glowinski84}.
This method is called the {\it projected SOR (PSOR) method} and it reads
\begin{equation}\label{PSOR}
U_i^{(k+1)} = \max\left\{ U_i^{(k)}
+ \frac{\omega}{\BB_{i,i}} \left( \Psi_i
- \sum_{j=1}^{i-1} \BB_{i,j}^{} U_j^{(k+1)}
- \sum_{j=i}^m \BB_{i,j}^{} U_j^{(k)} \right), \;
\Phi_i \right\}\quad
\end{equation}
$(i=1,2,\ldots,m,~ k=0,1,2,\ldots)$.
As can be expected, the PSOR method suffers from the same drawback
as the SOR method mentioned above.

\subsection{Multigrid Methods}
The aim of multigrid methods for solving linear systems \eqref{linsys} is to
render the number of iterations essentially independent of the problem size $m$.
The stationary iterative methods typically reduce high frequency errors 
quickly, while low frequency errors are reduced much more slowly. 
The idea of multigrid methods is to compute efficiently corrections to
these slowly varying errors on coarser spatial grids. 
The multigrid methods can be divided
into geometrical and algebraic methods. With the geometrical methods
discretizations are explicitly constructed on a sequence of grids
and transfer operators between these grids are explicitly defined.
Algebraic multigrid (AMG) methods \cite{Ruge87,Stueben01}
build the coarse problems and the transfer operators automatically using
the properties of the matrix $\BB$. The details of these methods are
beyond the scope of this chapter and we refer to e.g.~\cite{Trottenberg01}
for details and extensive literature on this.

Several versions of multigrid methods also exist for LCPs. Brandt and
Cryer introduced in \cite{Brandt83} a projected full approximation scheme
(PFAS) multigrid method for LCPs. American options under stochastic
volatility were priced using the PFAS method in \cite{Clarke99,Oosterlee03}.
A projected multigrid (PMG) method for LCPs introduced in \cite{Reisinger04}
resembles more closely a classical multigrid method for linear problems.
This method has been used to price American options in
\cite{Ikonen08,Reisinger04}. Recently, an AMG method was generalized
for LCPs in \cite{Toivanen12}. The resulting method is called
the projected algebraic multigrid (PAMG) method and
resembles the PMG method in the treatment of the complementarity conditions.

\section{Numerical Illustrations}\label{experiments}
\begin{figure}
\begin{center}
\includegraphics[width=0.7\columnwidth]{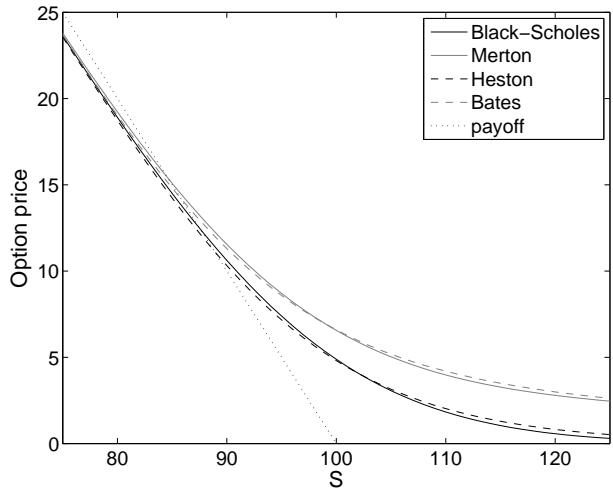}
\end{center}
\caption{The fair values of European put options for
the asset prices $75 \le s \le 125$ and the volatility $\sigma = 0.2$
(the variance $v = 0.04$) under the four considered models.}
\label{EuropeanPuts}
\end{figure}

\begin{figure}
\begin{center}
\includegraphics[width=0.7\columnwidth]{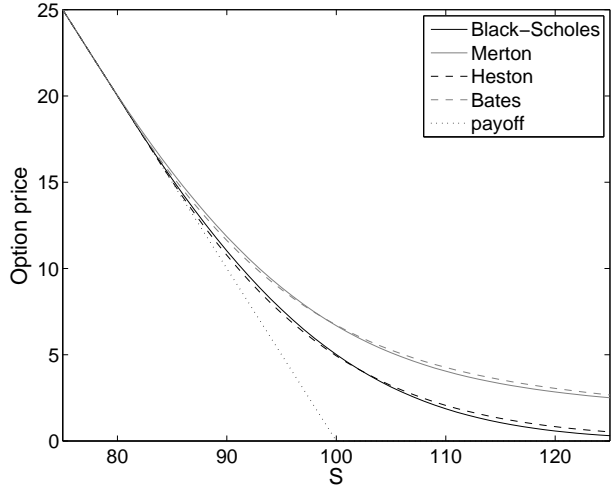}
\end{center}
\caption{The fair values of American put options for
the asset prices $75 \le s \le 125$ and the volatility $\sigma = 0.2$
(the variance $v = 0.04$) under the four considered models.}
\label{AmericanPuts}
\end{figure}

In the following we price European and American put options under a hierarchy
of models: Black--Scholes, Merton, Heston, and Bates. 
The interest rate, the maturity time, and the strike price
are always taken as
\begin{equation*}
r=0.03,~~T=0.5,~~\text{and}~~K=100.
\end{equation*}
For the purpose of illustration,
Fig.~\ref{EuropeanPuts} and Fig.~\ref{AmericanPuts} show 
fair values of European and American options, respectively, 
under the four considered models with the model parameters 
described in the following sections.

\subsection{Black--Scholes model}\label{BlackScholesExp}
In the case of the Black--Scholes model, we price American put options.  
The volatility in the model \eqref{GeomBrownSDE} is taken as
\begin{equation*}
\sigma=0.2
\end{equation*}
and the following boundary conditions are employed:
\begin{eqnarray}
u(0,t) = &K
&~~{\rm for}~~~0 < t\le T,\label{BSBC1}\\
u_s(S_{\max},t) = &0
&~~{\rm for}~~~0 < t\le T.\label{BSBC2}
\end{eqnarray}
The Neumann boundary condition \eqref{BSBC2} introduces a modeling error as
it is not exactly fulfilled by the actual option price function. If $S_{\max}$
is taken sufficiently large, however, this error will be small in the region
of interest.

For the spatial discretization of the Black--Scholes PDE \eqref{BSPDE}, 
we apply FD formulas on nonuniform grids such that a large fraction
of the grid points lie in the region of interest, that is, in the neighborhood
of $s=K$.

For the construction of the spatial grid we adopt \cite{Haentjens12}.
Let integer $m_1\ge 1$, constant $c>0$,
and $0<S_{\lleft}<K<S_{\rright}<S_{\max}$ be given. Let equidistant points
$\xi_{\min}=\xi_0 < \xi_1 < \ldots < \xi_{m_1}=\xi_{\max}$ be given with distance
$\Delta \xi$ and
\begin{align*}
\xi_{\min} &= \sinh^{-1}\left( \frac{- S_{\lleft}}{c} \right),\\
\xi_{\iin} &= \frac{S_{\rright}-S_{\lleft}}{c},\\
\xi_{\max} &= \xi_{\iin} + \sinh^{-1}\left( \frac{S_{\max} - S_{\rright}}{c} \right).
\end{align*}
Then we define a nonuniform grid $0=s_0 < s_1 < \ldots < s_{m_1}=S_{\max}$ by the
transformation
\begin{equation}\label{sgrid}
s_i = \varphi(\xi_i) \quad (0\le i \le m_1),
\end{equation}
where
\begin{equation*}
\varphi(\xi) =
\begin{cases}
 S_{\lleft}\,\, + c\cdot\sinh(\xi) & ~(\xi_{\min} \leq \xi \le 0),\\
 S_{\lleft}\,\, + c\cdot\xi & ~(0 < \xi < \xi_{\iin} ),\\
 S_{\rright} + c\cdot\sinh(\xi-\xi_{\iin}) & ~(\xi_{\iin} \le \xi \leq \xi_{\max}).
\end{cases}
\end{equation*}
The grid \eqref{sgrid} is uniform inside $[S_{\lleft},S_{\rright}]$ and nonuniform
outside. The parameter $c$ controls the fraction of grid points $s_i$ that lie inside 
$[S_{\lleft},S_{\rright}]$.
The grid is smooth in the sense that there exist real constants $C_0, C_1, C_2 >0$
such that the grid sizes $\Delta s_i = s_i-s_{i-1}$ satisfy
\begin{equation}\label{smooth}
  C_0\, \Delta \xi \le \Delta s_i \le C_1\, \Delta \xi \quad {\rm and} \quad
  |\Delta s_{i+1} - \Delta s_i| \le C_2 \left( \Delta \xi \right)^2
\end{equation}
uniformly in $i$ and $m_1$.
For the parameters in the grid we make the (heuristic) 
choice
\begin{equation*}
S_{\max} = 8K,~~
c = \frac{K}{10},~~
S_{\lleft}=\max \left(\tfrac{1}{2},e^{-T/10}\right)K~,~~
S_{\rright}=\min \left(\tfrac{3}{2},e^{T/10}\right)K.
\end{equation*}

The semidiscretization of the initial-boundary value problem for the
Black--Scholes PDE is then performed as follows. At the interior
grid points each spatial derivative appearing in \eqref{BSPDE} is
replaced by its corresponding second-order central FD
formula described in Section \ref{spacediscr}. At the boundary $s=S_{\max}$
the Neumann condition \eqref{BSBC2} gives $\partial u/ \partial s$.
Next, $\partial^2 u/ \partial s^2$ is approximated by the central formula
with the value at the virtual point $S_{\max}+\Delta s_{m_1}$ 
defined by linear extrapolation using \eqref{BSBC2}.

Concerning the initial condition, we always replace the value 
of the payoff function $\phi$ at the grid point $s_i$ nearest to the strike $K$
by its cell average,
\begin{equation*}
\frac{1}{h}
\int_{s_{i-1/2}}^{s_{i+1/2}} \max(K-s,0)\, ds,
\end{equation*}
where
\begin{equation*}
s_{i-1/2} = \tfrac{1}{2}(s_{i-1}+s_i),~~
s_{i+1/2} = \tfrac{1}{2}(s_i+s_{i+1}),~~
h = s_{i+1/2}-s_{i-1/2}.
\end{equation*}
This reduces the dependency of the discretization error on the location of
the strike relative to the $s$-grid, see e.g.~\cite{Tavella00}.

The time discretization is performed by the Crank--Nicolson method 
with Rannacher smoothing. The time stepping is started by taking two
backward Euler steps using the time step $\tfrac{1}{2} \Delta t$. With
this choice all time steps are performed with the same coefficient
matrix $\II - \tfrac{1}{2} \Delta t \AA$. Furthermore, halving the time
step with the Euler method helps to reduce the additional error caused by
this method. 
Note that we count these two Euler steps as one time step in order to keep 
the notations convenient.

\begin{figure}
\begin{center}
\includegraphics[width=0.7\columnwidth]{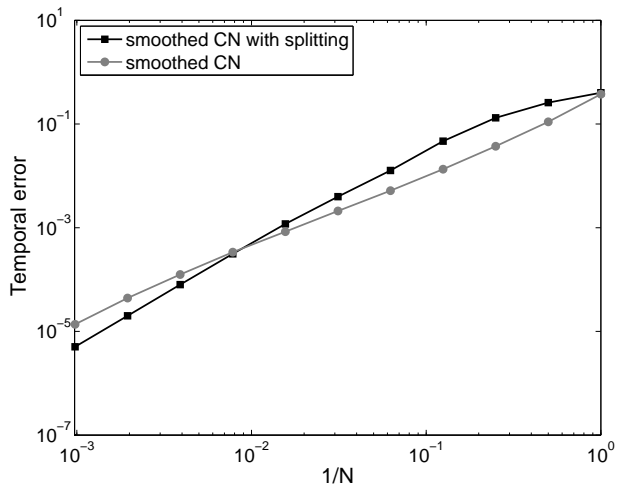}
\end{center}
\caption{
The temporal discretization errors for the American option under the
Black--Scholes model for the smoothed Crank--Nicolson method with and 
without the operator splitting method for LCPs.}
\label{BlackScholesAmerTimeError}
\end{figure}

We define the {\it temporal discretization error} to be
\begin{equation}\label{temperror1d}
\widehat{e} (m_1,N) = \max \left\{\, | U_{N,i} - U_i(T) |:~
\tfrac{1}{2} K < s_i < \tfrac{3}{2} K \right\},
\end{equation}
where $U_{N,i}$ denotes the component of the vector $U_N$ associated
to the grid point~$s_i$. 
We study the temporal discretization errors on the
grids $(m_1,N) = (160,2^k)$ for $k=0,1,\ldots,10$.
The reference price vector $U(T)$ is computed
using the space-time grid $(160,5000)$.
Fig.~\ref{BlackScholesAmerTimeError} compares the temporal errors 
of the smoothed Crank--Nicolson method with and without 
the operator splitting method for LCPs described in Section \ref{LCPsplit}.
For larger time steps the Crank--Nicolson method without splitting
is more accurate. In this example the convergence rate of the 
splitted method is slightly less than second-order and a 
bit higher than the convergence rate of the unsplitted method. 
Thus, for smaller time steps the operator splitting method is slightly
more accurate.

\subsection{Merton model}\label{MertonExp}
Under the Merton jump diffusion model, we price European and American put options.
For the jump part of the model, the jump activity, the mean of the normal distribution,
and its standard deviation are taken as
\begin{equation}\label{MertonParam}
\lambda = 0.2,~~\delta = 0.4,~~\text{and}~~\gamma = -0.5,
\end{equation}
respectively; see \eqref{MertonPDF}.
The boundary condition at $s = 0$ is given by \eqref{BSBC} for
the European put option and by \eqref{BSBC1} for the American put option.
At the truncation boundary $s = S_{\max}$, we use the Neumann boundary
condition \eqref{BSBC2}.

The same space-time grids are considered as with the Black--Scholes model 
in Section
\ref{BlackScholesExp} and also the spatial derivatives are discretized
in the same way. For the integral term, we use a linear interpolation for $u$
between grid points and take $u$ to be zero for $s > S_{\max}$. The formulas
for the resulting matrix $\JJ$ are given in \cite{Salmi11}, for example.

For the time discretization, we apply the IMEX--CNAB scheme, 
which is always smoothed by
two Euler steps with the time step $\tfrac{1}{2}\Delta t$. In these first
steps the backward Euler method is used for the discretized differential
part $\DD$ and the forward Euler method is used for the 
discretized integral part $\JJ$. For European options, these steps are given by
\begin{equation*}
\begin{split}
\left( \II - \tfrac{\Delta t}{2} \DD \right) U_{1/2} &= U_0 +
\tfrac{\Delta t}{2} \JJ U_0 + \tfrac{\Delta t}{2} G_{1/2}, \\
\left( \II - \tfrac{\Delta t}{2} \DD \right) U_1 &= U_{1/2} +
\tfrac{\Delta t}{2} \JJ U_{1/2} + \tfrac{\Delta t}{2} G_1.
\end{split}
\end{equation*}
In the absence of jumps, these steps reduce to the same Rannacher
smoothing used with the Black--Scholes model.
After these two steps the IMEX--CNAB scheme defined by \eqref{IMEXCNAB}
is employed.

\begin{figure}
\begin{center}
\includegraphics[width=0.7\columnwidth]{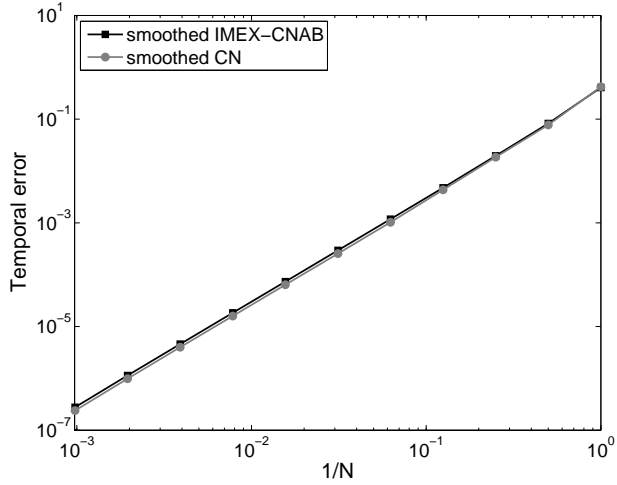}
\end{center}
\caption{The temporal discretization errors for
the European option under the Merton model with the IMEX--CNAB scheme
and the Crank--Nicolson method, both with smoothing.}
\label{MertonTimeError}
\end{figure}

We study the temporal discretization errors for European and American options
on the same grids $(m_1,N) = (160,2^k)$, $k=0,1,\ldots,10$, and using the
same error measure \eqref{temperror1d} as before. Fig.~\ref{MertonTimeError}
shows the temporal errors for the European option using the IMEX--CNAB scheme
and the Crank--Nicolson method with classical Rannacher smoothing. 
We observe that the temporal errors for the two methods are essentially
the same and they exhibit second-order convergence.

\begin{figure}
\begin{center}
\includegraphics[width=0.7\columnwidth]{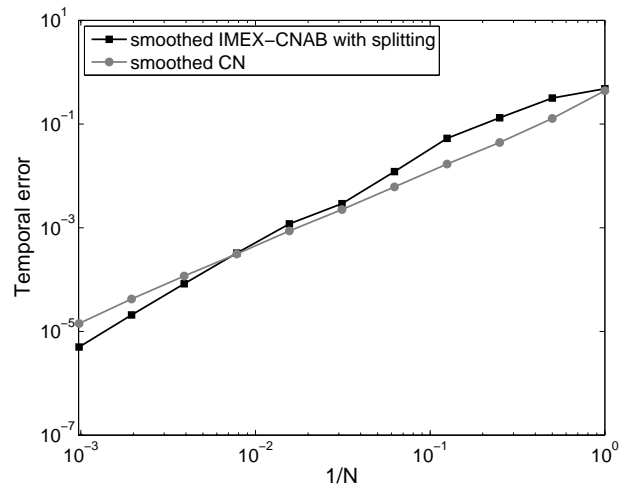}
\end{center}
\caption{The temporal discretization errors for the American option
under the Merton model with the IMEX--CNAB scheme
together with the operator splitting method for LCPs, and
the Crank--Nicolson method, both with smoothing.}
\label{MertonAmerError}
\end{figure}

Fig.~\ref{MertonAmerError} shows the same temporal errors for American
options using the IMEX--CNAB scheme with operator splitting
for LCPs and the Crank--Nicolson method without splitting.
The convergence result for the two methods is very similar 
to the case of the Black--Scholes model in Section \ref{BlackScholesExp}. 
Thus, for larger time steps the Crank--Nicolson method is 
more accurate while for smaller time steps the 
IMEX--CNAB scheme with splitting is more accurate.

\begin{figure}
\begin{center}
\includegraphics[width=0.7\columnwidth]{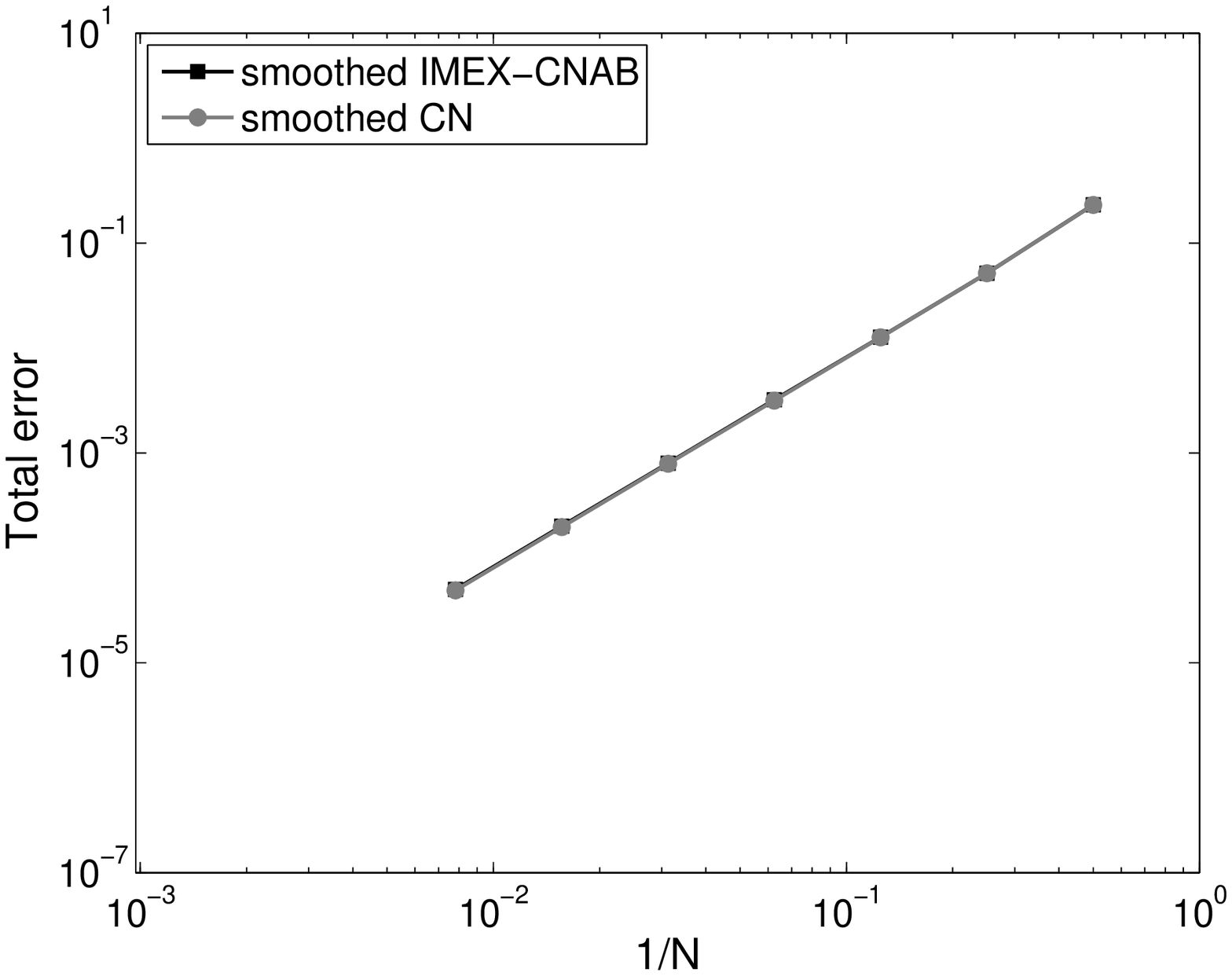}
\end{center}
\caption{The total discretization errors for
the European option under the Merton model with the IMEX--CNAB scheme
and the Crank--Nicolson method, both with smoothing.}
\label{MertonError}
\end{figure}

In order to gauge the effectiveness of the proposed discretizations, we
report the total discretization errors for the European option on the
space-time refining grids $(m_1,N) = 2^k (10,2)$, $k=0,1,\ldots,6$.
The {\it total discretization error} is defined by
\begin{equation}\label{totalerror1d}
e (m_1,N) = \max \left\{\, | U_{N,i} - u(s_i,T) |:~
\tfrac{1}{2} K < s_i < \tfrac{3}{2} K \right\}.
\end{equation}
The reference price function $u$ is computed on the space-time grid
$(10240,2048)$. Fig.~\ref{MertonError} shows the total error for
the European option using the IMEX--CNAB scheme and the Crank--Nicolson
method. As with the temporal 
errors the total errors for both methods are essentially the same and
both show a second-order convergence behavior.

\subsection{Heston model}\label{HestonExp}
Under the Heston stochastic volatility model we consider European and American put options as well. 
For the mean-reversion rate, the long-term mean, the volatility-of-variance and the correlation the following values are taken:
\begin{equation}\label{HestonParam}
\kappa = 2,~~\eta=0.04,~~\sigma=0.25,~~\text{and}~~
\rho=-0.5.
\end{equation}
The spatial domain is truncated to $[0,S_{\max}]\times [0,V_{\max}]$ with $S_{\max}=8K$ and $V_{\max}=5$.
The following boundary conditions are imposed:
\begin{eqnarray}
u(0,v,t) = &d\!f\cdot K
&~~{\rm for}~~~0\le v \le V_{\max},~0 < t\le T,\label{HBC1}\\
u_s(S_{\max},v,t) = &0
&~~{\rm for}~~~0\le v \le V_{\max},~0 < t\le T,\label{HBC2}\\
u_v(s,V_{\max},t) = &0 
&~~{\rm for}~~~0\le s \le S_{\max},~0 < t\le T,\label{HBC3}
\end{eqnarray}
where $d\!f= e^{-rt}$ in the European case and $d\!f=1$ in the American case.
At the degenerate boundary $v=0$ the Heston PDE holds in the European case and it is assumed that the Heston LCP holds in the American case.
The two conditions at $s=S_{\max}$ and $v=V_{\max}$ introduce a modeling
error, as they are not exactly fulfilled by the actual option price function,
but in our experiments this error is small on the region of interest in
the $(s,v)$-domain.

For the spatial discretization of the Heston PDE and Heston LCP we apply FD formulas on Cartesian grids.
Here nonuniform grids are used in both the $s$- and $v$-directions such that a large fraction of the grid points lie in the neighborhoods of $s=K$ and $v=0$, respectively.
This is the region in the $(s,v)$-domain where one wishes to obtain option prices.
Next, the application of such nonuniform grids can greatly improve the accuracy of the FD discretization as compared to using uniform grids.
This is related to the facts that the initial function (\ref{payoff}) possesses a discontinuity in its first derivative at $s=K$ and that for $v\approx 0$ the Heston PDE is convection-dominated.
The grid in the $s$-direction is taken identical to that in Section \ref{BlackScholesExp}.

To construct the grid in the $v$-direction, let integer $m_2\ge 1$ and constant $d>0$ and let equidistant points be given by $\psi_j = j\cdot \Delta \psi$ 
for $j=0,1,\ldots,m_2$ with
\begin{equation*}
\Delta \psi = \frac{1}{m_2}\, \sinh^{-1} \left(\frac{V_{\max}}{d}\right).
\end{equation*}
Then a smooth, nonuniform grid $0=v_0 < v_1 < \ldots < v_{m_2}=V_{\max}$
is defined by
\begin{equation}\label{v-mesh}
v_j = d\cdot{\rm sinh}(\psi_j) \quad (0\le j \le m_2).
\end{equation}
The parameter $d$ controls the fraction of grid points $v_j$
that lie near $v=0$. We heuristically choose
\begin{equation*}
d=\frac{V_{\max}}{500}\,.
\end{equation*}

The semidiscretization of the initial-boundary value problem for the Heston PDE and Heston LCP is performed as follows.
In view of the Dirichlet condition (\ref{HBC1}), the grid in $[0,S_{\max}]\times [0,V_{\max}]$ is given by $\{ (s_i,v_j): 1\le i \le m_1,~0\le j \le m_2 \}.$
At this grid, each spatial derivative is replaced by its corresponding second-order central FD formula described in Section \ref{spacediscr} with a modification for the boundaries $v=0$, $s=S_{\max}$,
and $v=V_{\max}$.

At the boundary $v=0$ the derivative $\partial u/ \partial v$ is approximated using a second-order forward formula.
All other derivative terms in the $v$-direction vanish at $v=0$, and therefore do not require further treatment.

At the boundary $s=S_{\max}$ the spatial derivatives in the $s$-direction are dealt with as in Section \ref{BlackScholesExp}.
Note that the Neumann condition (\ref{HBC2}) at $s=S_{\max}$ implies that the mixed derivative $\partial^2 u / \partial s \partial v $ vanishes there.

At the boundary $v=V_{\max}$ the spatial derivatives in the $v$-direction need to be considered. This is done fully analogously to those in the $s$-direction at $s=S_{\max}$ using now the Neumann condition (\ref{HBC3}). 

Define the temporal discretization error by
\begin{equation}\label{temperror2d}
\widehat{e} (m_1,m_2,N) = \max \left\{\, | U_{N,l} - U_l(T) |:~
\tfrac{1}{2} K < s_i < \tfrac{3}{2} K,~0< v_j <1 \right\},
\end{equation}
where the index $l$ corresponds to the grid point $(s_i,v_j)$. 
The reference vector $U(T)$ is computed using $(m_1,m_2,N) =(160,80,5000)$.
We study these errors for $(m_1,m_2,N) = (160,80, 2^k)$ with $k=0,1,\ldots,10$ and three methods:
the Do scheme with $\theta = \frac{1}{2}$ and smoothing, the MCS scheme with $\theta = \frac{1}{3}$ without smoothing, and the Crank--Nicolson scheme with smoothing.

\begin{figure}
\begin{center}
\includegraphics[width=0.7\columnwidth]{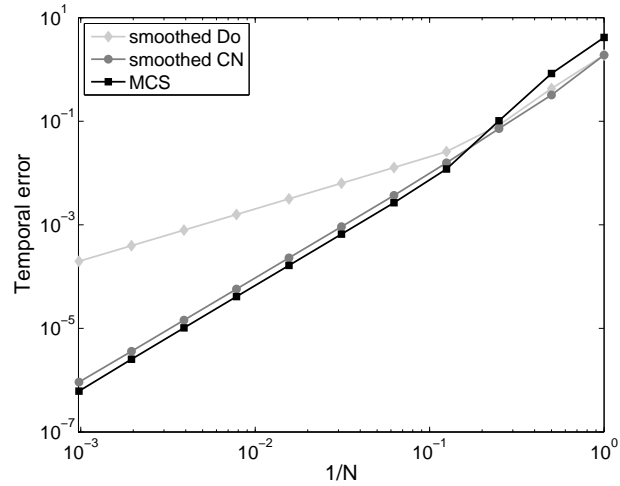}
\end{center}
\caption{Temporal discretization errors in the case of the European put
option under the Heston model. The time discretization methods are:
the Do scheme with $\theta = \frac{1}{2}$ and smoothing,
the MCS scheme with $\theta = \frac{1}{3}$ without smoothing,
and the Crank--Nicolson scheme with smoothing.}
\label{HestonTimeError}
\end{figure}

Fig.~\ref{HestonTimeError} displays the obtained results for the European put option.
As a first observation, for all three methods the temporal errors are bounded from above by a moderate value and decrease monotonically as $N$ increases.
The error graphs for the MCS and Crank--Nicolson schemes are almost identical and reveal a second-order convergence behavior.
The Do scheme only shows first-order convergence.
Clearly, the convergence orders observed for the three methods agree with their respective classical orders of consistency.
Additional experiments by substantially changing $(m_1,m_2)$ indicate that for all three methods the temporal errors are almost unaffected, which is a desirable property and suggests convergence in the so-called {\it stiff sense}.
Whereas their results are not displayed, we mention that the CS scheme with $\theta = \frac{1}{2}$ and smoothing and the HV scheme with $\theta = \frac{1}{2}+\frac{1}{6}\sqrt{3}$ without smoothing behave similarly to the MCS scheme in this experiment, with slightly larger errors.

\begin{figure}
\begin{center}
\includegraphics[width=0.7\columnwidth]{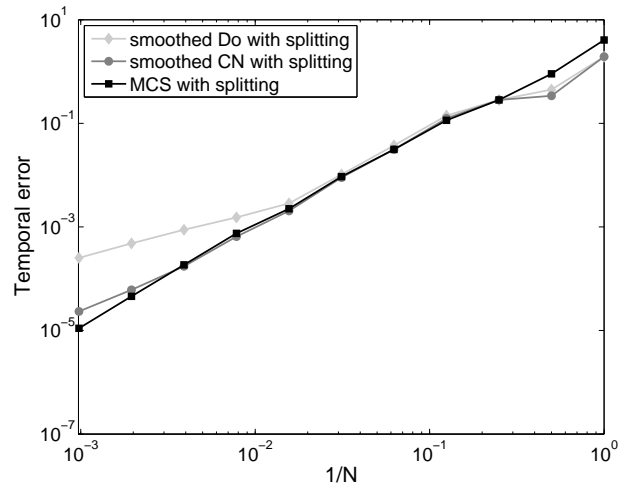}
\end{center}
\caption{Temporal discretization errors in the case of the American put
option under the Heston model. The time discretization methods are:
the Do scheme with $\theta = \frac{1}{2}$ and smoothing,
the MCS scheme with $\theta = \frac{1}{3}$ without smoothing,
and the Crank--Nicolson scheme with smoothing.}
\label{HestonAmerError}
\end{figure}

Fig.~\ref{HestonAmerError} displays the obtained results for the American put option.
Our observations are analogous to those made above in the case of the European option.
It is interesting to note, however, that the Do scheme often has temporal errors that are almost the same as for the MCS and Crank--Nicolson schemes.
But if $N$ gets sufficiently large, then a first-order convergence behavior for this method indeed sets in.
For the Crank--Nicolson scheme a small deviation from second-order is seen when $N$ is large.
This disappears however when other values $(m_1,m_2)$ are considered.
Additional experiments by substantially changing $(m_1,m_2)$ indicate that for all three methods the temporal errors are at most mildly affected.

We next consider, in the European put option case, the total discretization error defined by
\begin{equation}\label{totalerror2d}
e (m_1,m_2,N) = \max \left\{\, | U_{N,l} - u(s_i,v_j,T) |:~
\tfrac{1}{2} K < s_i < \tfrac{3}{2} K,~0< v_j <1 \right\},
\end{equation}
with index $l$ corresponding to the grid point $(s_i,v_j)$. 
Here exact solution values $u$ are computed by a suitable implementation of Heston's semi-closed form analytical formula \cite{Heston93}.
Note that the modeling error, which is due to the truncation of the domain of the Heston PDE to a bounded set, is also contained in $e(m_1,m_2,N)$.
In our experiment, this contribution is negligible.

\begin{figure}
\begin{center}
\includegraphics[width=0.7\columnwidth]{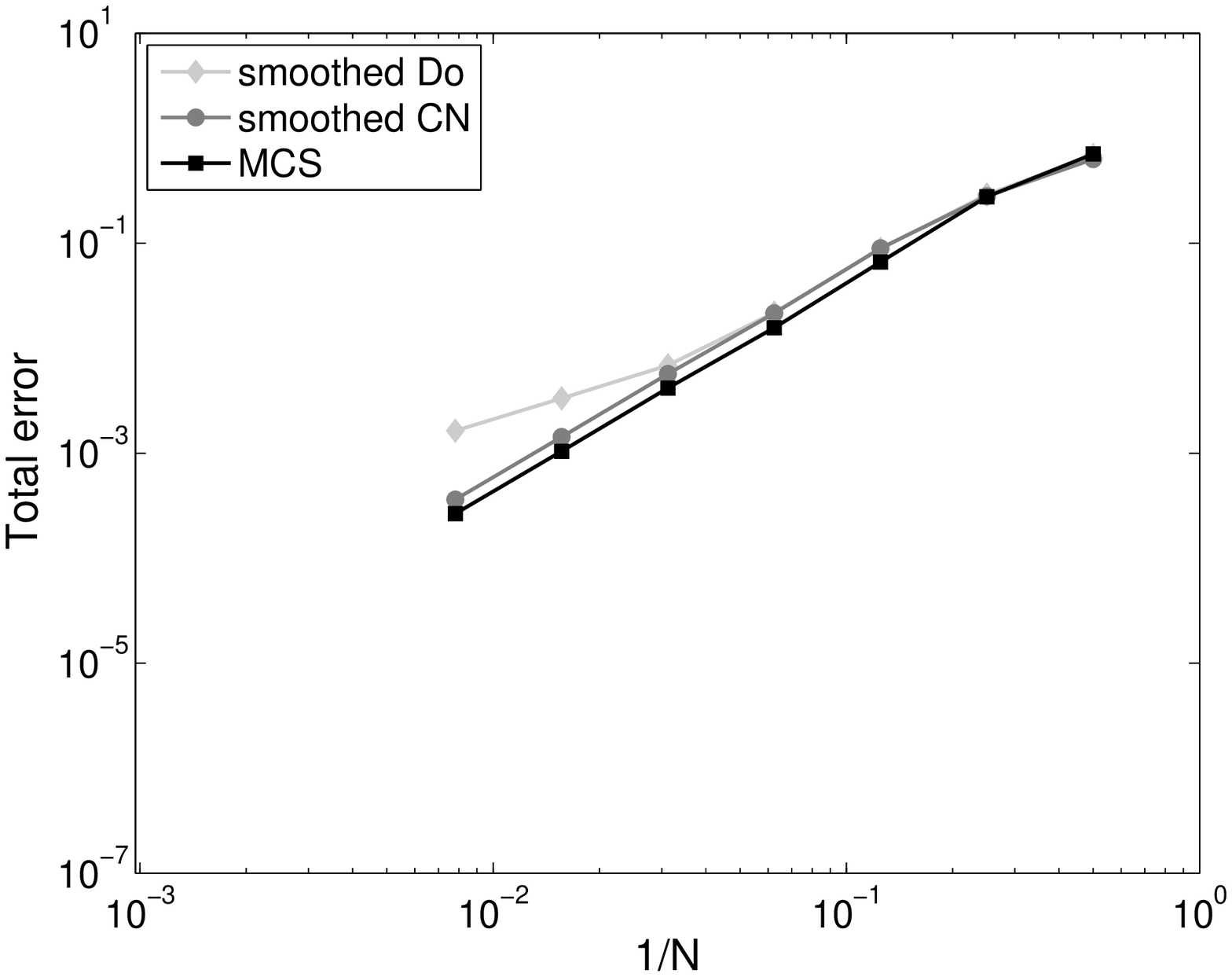}
\end{center}
\caption{Total discretization errors in the case of the European put
option under the Heston model. The time discretization methods are:
the Do scheme with $\theta = \frac{1}{2}$ and smoothing,
the MCS scheme with $\theta = \frac{1}{3}$ without smoothing,
and the Crank--Nicolson scheme with smoothing.}
\label{HestonError}
\end{figure}

Fig.~\ref{HestonError} displays the total discretization errors for $(m_1,m_2,N) = 2^k (10,5,2)$ with $k=0,1,\ldots,6$ and the three schemes under consideration in this section.
With the MCS and Crank--Nicolson schemes the total errors are essentially the same and a second-order convergence behavior is observed.
With the Do scheme, the total errors are almost same as these two schemes up to $k=4$, but then the convergence drops to the expected first-order.

For a more extensive numerical study of ADI schemes in the (two-dimensional) Heston model we refer to \cite{intHout10} for European-style options and to \cite{Haentjens15} for American-style options.
For three-dimensional PDEs in finance, such as the HHW PDE, the numerical convergence of ADI schemes has been investigated in \cite{Haentjens13a,Haentjens12} and for a four-dimensional PDE in \cite{Haentjens13c}.
In these references a variety of parameter sets has been considered, including long maturity times and cases where the Feller condition is strongly violated, together with various barrier options and the approximation of hedging quantities.

\subsection{Bates model}
We price European and American put options under the Bates model.
The boundary conditions are given by \eqref{HBC1}--\eqref{HBC3}.
For the stochastic volatility part of the model the parameters are taken
the same as for the Heston model and they are given by \eqref{HestonParam}.
For the jump part, the parameters are the same as for the Merton model
and they are given by \eqref{MertonParam}. The discretizations are
based on the same grids and the spatial derivatives are discretized
in the same way as with the Heston model in Section \ref{HestonExp}.
For the jump integral, the same discretization
is used as with the Merton model in Section \ref{MertonExp}.
We consider here the IMEX--CNAB scheme and Crank--Nicolson 
method both applied with smoothing as for the Merton model.

\begin{figure}
\begin{center}
\includegraphics[width=0.7\columnwidth]{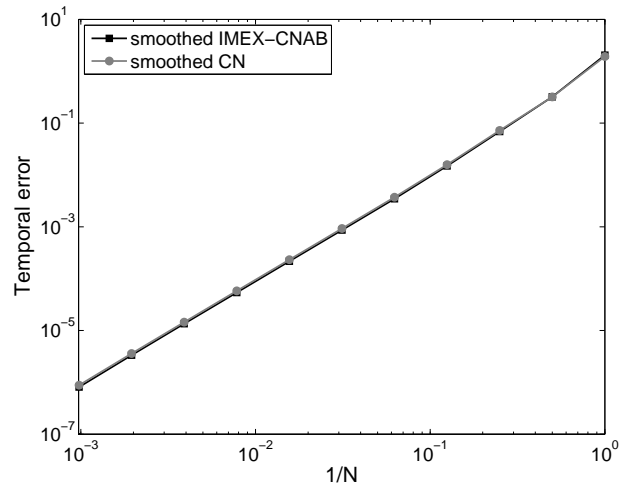}
\end{center}
\caption{The temporal discretization errors for
the European option under the Bates model with the IMEX--CNAB scheme
and the Crank--Nicolson method, both with smoothing.}
\label{BatesTimeError}
\end{figure}

\begin{figure}
\begin{center}
\includegraphics[width=0.7\columnwidth]{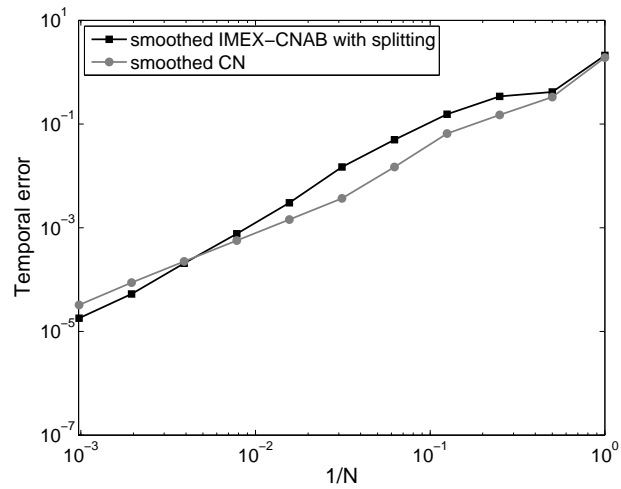}
\end{center}
\caption{The temporal discretization errors for
the American option under the Bates model with the IMEX--CNAB scheme
together with the operator splitting method for LCPs, and the
Crank--Nicolson method, both with smoothing.}
\label{BatesAmerTimeError}
\end{figure}

As with the Heston model, we consider the temporal discretization
errors on the grids $(m_1,m_2,N) = (160,80,2^k)$, $k=0,1,\ldots,10$.
The reference price vector $U(T)$ is computed using the space-time grid
$(160,80,5000)$. The temporal discretization errors $\widehat{e} (m_1,m_2,N)$
are shown for the European option in Fig.~\ref{BatesTimeError} 
and for the American option in Fig.~\ref{BatesAmerTimeError}. 
The plots show the errors for the IMEX--CNAB scheme and the Crank--Nicolson 
method. For the American option the 
operator splitting method for LCPs is used with the IMEX--CNAB scheme.
As with other models, the temporal errors for the European option
are very similar for both methods and they both exhibit second-order
convergence. For the American option, the difference between the
methods is less pronounced than with the Black--Scholes and Merton models.
Still the Crank--Nicolson method is slightly more 
accurate than the operator splitting method for large time 
steps and the reverse is true for small time steps. In this example the 
convergence rates seem to be between 1.5 and 2.0.

\begin{figure}
\begin{center}
\includegraphics[width=0.7\columnwidth]{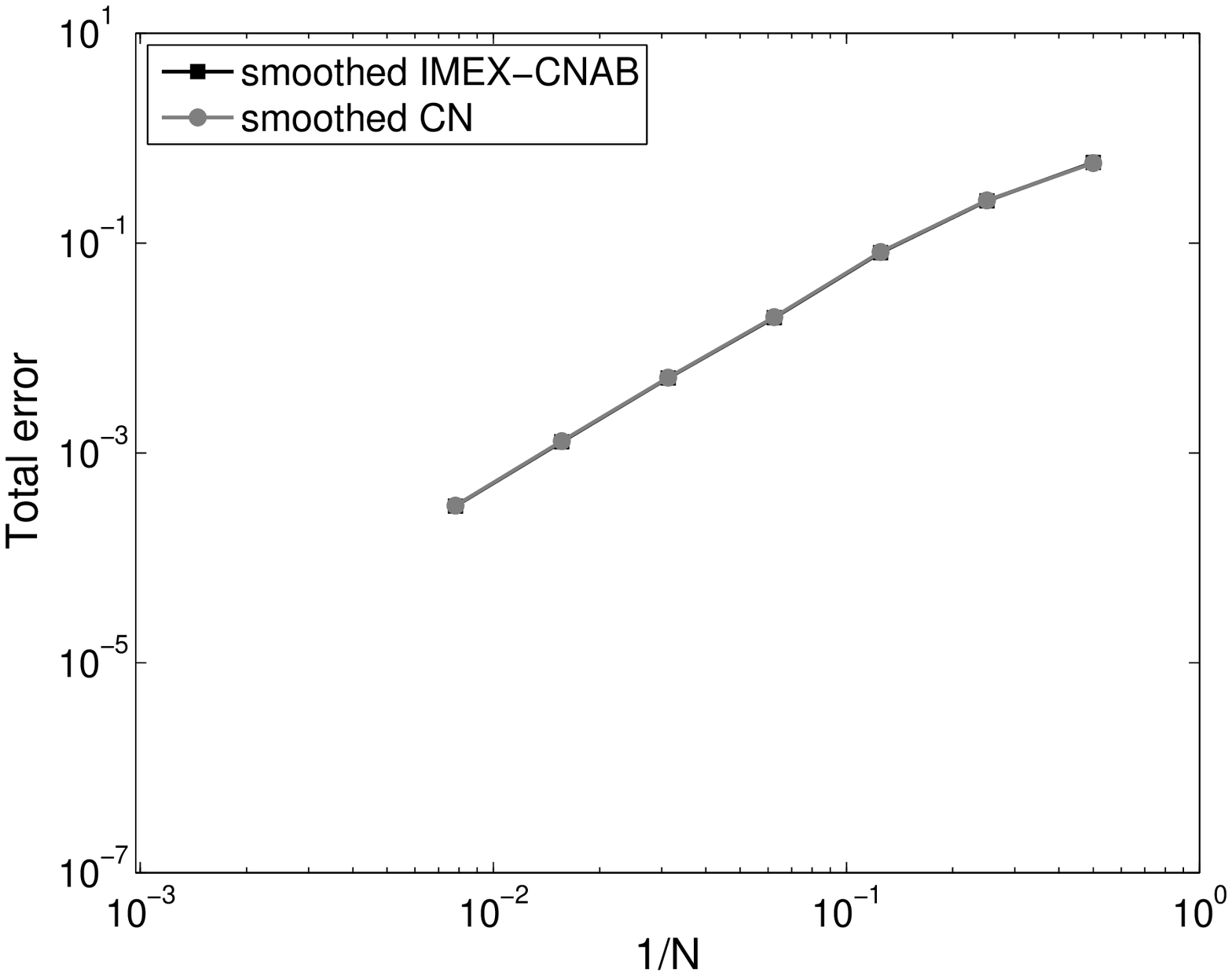}
\end{center}
\caption{The total discretization errors for
the European option under the Bates model with the IMEX--CNAB scheme
and the Crank--Nicolson method, both with smoothing.}
\label{BatesError}
\end{figure}

We computed the total discretization errors $e(m_1,m_2,N)$ for
the European option on the grids $(m_1,m_2,N) = 2^k (10,5,2)$,
$k=0,1,\ldots,6$. The reference prices are computed on the
space-time grid $(2560,1280,512)$. Fig.~\ref{BatesError} shows
the total errors for the IMEX--CNAB scheme and the Crank--Nicolson
method. As with the other models,
the total errors for both methods are virtually the same and
both have second-order convergence of the total error.

\section{Conclusions}
We have discussed numerical solution methods for financial option valuation 
problems in the contemporary partial(-integro) differential equations framework. 
These problems are often multidimensional and can involve nonlocal 
integral operators due to jumps incorporated in the underlying asset 
price models. 
The early exercise feature of American-style options gives rise to linear 
complementarity problems, which are nonlinear.
All these properties add complexity to the discrete problems obtained by 
classical implicit numerical methods and renders their efficient solution
a challenging task.
The efficient computation of option values is, however, crucial 
in many applications.
In this chapter an overview has been given of various types of operator splitting 
methods for the discretization in time, which yield in each time step a sequence 
of discrete subproblems that can be handled much more easily and efficiently 
without essentially influencing the accuracy of the underlying discretization.
The following highlights the different operator splitting methods presented in 
this chapter.

For multidimensional models the directional splitting methods of the ADI type 
offer a fast, accurate, and easy-to-implement way for the numerical time stepping.
They are adapted to effectively deal with mixed spatial derivative terms, which 
are ubiquitous in finance.
ADI schemes lead to a sequence of sparse linear subproblems that can be solved
by LU decomposition at optimal computational cost, that is, the number of required 
operations is directly proportional to the number of unknowns.
The MCS and HV schemes, with a proper choice of their parameter $\theta$, are 
recommended as these show stability and second-order convergence and reveal a
better inherent smoothing than second-order CS.

The spatial discretization of jumps models for the underlying asset price
yields dense matrices. All classical implicit time discretization schemes
require solving systems with these dense matrices. By employing
an IMEX method like the IMEX--CNAB scheme advocate here, with an
explicit treatment of (finite activity) jumps and an implicit treatment of
the remainder of the operator, each time step involves only multiplications
with these dense matrices. This is computationally a much easier task and
can be often performed very fast using FFT. The accuracy and stability
of the IMEX--CNAB scheme are good when the jump activity is not very high, 
e.g. less than several jumps per year.

Iterative methods like the PSOR method for solving LCPs resulting from the
pricing of American-style options often converge slowly. 
We discussed an operator splitting method based on a Lagrange multiplier 
formulation, treating in each time step the early exercise constraint and 
complementarity condition in separate subproblems, where the main subproblem 
is essentially the same as for the European-style counterpart. With this 
approach it is easy to adapt a European option pricer to American options. 
We presented such an adaptation for ADI and IMEX methods. 
Also, it is applicable for most models of underlying asset prices. 
Numerical experience with this operator splitting method indicates
that the accuracy stays essentially the same as in the case of the original 
LCP, but there can be a major reduction in computational time.

\bibliographystyle{spmpsci}
\bibliography{paper}

\begin{thebibliography}{10}
\providecommand{\url}[1]{{#1}}
\providecommand{\urlprefix}{URL }
\expandafter\ifx\csname urlstyle\endcsname\relax
  \providecommand{\doi}[1]{DOI~\discretionary{}{}{}#1}\else
  \providecommand{\doi}{DOI~\discretionary{}{}{}\begingroup
  \urlstyle{rm}\Url}\fi

\bibitem{Achdou05}
Achdou, Y., Pironneau, O.: Computational methods for option pricing,
  \emph{Frontiers in Applied Mathematics}, vol.~30.
\newblock Society for Industrial and Applied Mathematics (SIAM), Philadelphia,
  PA (2005)

\bibitem{Almendral05}
Almendral, A., Oosterlee, C.W.: Numerical valuation of options with jumps in
  the underlying.
\newblock Appl. Numer. Math. \textbf{53}(1), 1--18 (2005)

\bibitem{Andersen00}
Andersen, L., Andreasen, J.: Jump-diffusion processes: Volatility smile fitting
  and numerical methods for option pricing.
\newblock Rev. Deriv. Res. \textbf{4}(3), 231--262 (2000)

\bibitem{Andersen10}
Andersen, L.B.G., Piterbarg, V.V.: Interest rate modeling, volume I:
  foundations and vanilla models.
\newblock Atlantic Financial Press (2010)

\bibitem{BIS13}
{Bank for International Settlements}: Triennial Central Bank Survey, Foreign
  exchange turnover in April 2013: preliminary global results (2013)

\bibitem{Bates96}
Bates, D.S.: Jumps and stochastic volatility: Exchange rate processes implicit
  in {D}eutsche {M}ark options.
\newblock Review Financial Stud. \textbf{9}(1), 69--107 (1996)

\bibitem{Black73}
Black, F., Scholes, M.: The pricing of options and corporate liabilities.
\newblock J. Political Economy \textbf{81}, 637--654 (1973)

\bibitem{Brandt83}
Brandt, A., Cryer, C.W.: Multigrid algorithms for the solution of linear
  complementarity problems arising from free boundary problems.
\newblock SIAM J. Sci. Statist. Comput. \textbf{4}(4), 655--684 (1983)

\bibitem{Brennan77}
Brennan, M.J., Schwartz, E.S.: The valuation of {A}merican put options.
\newblock J. Finance \textbf{32}, 449--462 (1977)

\bibitem{Briani07}
Briani, M., Natalini, R., Russo, G.: Implicit-explicit numerical schemes for
  jump-diffusion processes.
\newblock Calcolo \textbf{44}(1), 33--57 (2007)

\bibitem{Carr02}
Carr, P., Geman, H., Madan, D.B., Yor, M.: The fine structure of asset returns:
  an empirical investigation.
\newblock J. Business \textbf{75}, 305--332 (2002)

\bibitem{Carr07}
Carr, P., Mayo, A.: On the numerical evaluation of option prices in jump
  diffusion processes.
\newblock Eur. J. Finance \textbf{13}, 353--372 (2007)

\bibitem{Chorin68}
Chorin, A.J.: Numerical solution of the {N}avier-{S}tokes equations.
\newblock Math. Comp. \textbf{22}, 745--762 (1968)

\bibitem{Clark11}
Clark, I.J.: Foreign exchange option pricing.
\newblock Wiley, Chichester (2011)

\bibitem{Clarke99}
Clarke, N., Parrott, K.: Multigrid for {A}merican option pricing with
  stochastic volatility.
\newblock Appl. Math. Finance \textbf{6}, 177--195 (1999)

\bibitem{Cont04}
Cont, R., Tankov, P.: Financial modelling with jump processes.
\newblock Chapman \& Hall/CRC, Boca Raton, FL (2004)

\bibitem{Cont05}
Cont, R., Voltchkova, E.: A finite difference scheme for option pricing in jump
  diffusion and exponential {L}\'evy models.
\newblock SIAM J. Numer. Anal. \textbf{43}(4), 1596--1626 (2005)

\bibitem{Craig88}
Craig, I.J.D., Sneyd, A.D.: An alternating-direction implicit scheme for
  parabolic equations with mixed derivatives.
\newblock Comput. Math. Appl. \textbf{16}, 341--350 (1988)

\bibitem{Cryer71}
Cryer, C.W.: The solution of a quadratic programming problem using systematic
  overrelaxation.
\newblock SIAM J. Control \textbf{9}, 385--392 (1971)

\bibitem{Dang10}
Dang, D.M., Christara, C.C., Jackson, K.R., Lakhany, A.: A {PDE} pricing
  framework for cross-currency interest rate derivatives.
\newblock Proc. Comp. Sc. \textbf{1}, 2371--2380 (2010)

\bibitem{Davis06}
Davis, T.A.: Direct methods for sparse linear systems, \emph{Fundamentals of
  Algorithms}, vol.~2.
\newblock Society for Industrial and Applied Mathematics (SIAM), Philadelphia,
  PA (2006)

\bibitem{dHalluin05}
d'Halluin, Y., Forsyth, P.A., Vetzal, K.R.: Robust numerical methods for
  contingent claims under jump diffusion processes.
\newblock IMA J. Numer. Anal. \textbf{25}(1), 87--112 (2005)

\bibitem{Douglas56}
Douglas, J., Rachford, H.H.: On the numerical solution of heat conduction
  problems in two and three space variables.
\newblock Trans. Amer. Math. Soc. \textbf{82}, 421--439 (1956)

\bibitem{Egloff11}
Egloff, D.: {GPU}s in financial computing part {III}: {ADI} solvers on {GPU}s
  with application to stochastic volatility.
\newblock Wilmott Magazine (March), 51--53 (2011)

\bibitem{Ekstrom10}
Ekstr\"{o}m, E., Tysk, J.: The {B}lack--{S}choles equation in stochastic
  volatility models.
\newblock J. Math. Anal. Appl. \textbf{368}, 498–--507 (2010)

\bibitem{Feng08}
Feng, L., Linetsky, V.: Pricing options in jump-diffusion models: an
  extrapolation approach.
\newblock Oper. Res. \textbf{56}(2), 304--325 (2008)

\bibitem{Forsyth02}
Forsyth, P.A., Vetzal, K.R.: Quadratic convergence for valuing {A}merican
  options using a penalty method.
\newblock SIAM J. Sci. Comput. \textbf{23}(6), 2095--2122 (2002)

\bibitem{Frank97}
Frank, J., Hundsdorfer, W., Verwer, J.G.: On the stability of implicit-explicit
  linear multistep methods.
\newblock Appl. Numer. Math. \textbf{25}(2-3), 193--205 (1997)

\bibitem{George73}
George, A.: Nested dissection of a regular finite element mesh.
\newblock SIAM J. Numer. Anal. \textbf{10}, 345--363 (1973).
\newblock Collection of articles dedicated to the memory of George E. Forsythe

\bibitem{Glowinski84}
Glowinski, R.: Numerical methods for nonlinear variational problems.
\newblock Scientific Computation. Springer-Verlag, New York (1984)

\bibitem{Glowinski86}
Glowinski, R.: Splitting methods for the numerical solution of the
  incompressible {N}avier-{S}tokes equations.
\newblock In: Vistas in applied mathematics, Transl. Ser. Math. Engrg., pp.
  57--95. Optimization Software, New York (1986)

\bibitem{Grzelak11}
Grzelak, L.A., Oosterlee, C.W.: On the {H}eston model with stochastic interest
  rates.
\newblock SIAM J. Financial Math. \textbf{2}(1), 255--286 (2011)

\bibitem{Grzelak12}
Grzelak, L.A., Oosterlee, C.W., Van~Weeren, S.: Extension of stochastic
  volatility equity models with the {H}ull-{W}hite interest rate process.
\newblock Quant. Finance \textbf{12}(1), 89--105 (2012)

\bibitem{Haentjens13c}
Haentjens, T.: {ADI} schemes for the efficient and stable numerical pricing of
  financial options via multidimensional partial differential equations.
\newblock PhD thesis. University of Antwerp (2013)

\bibitem{Haentjens13a}
Haentjens, T.: Efficient and stable numerical solution of the
  {H}eston--{C}ox--{I}ngersoll--{R}oss partial differential equation by
  alternating direction implicit finite difference schemes.
\newblock Int. J. Comput. Math. \textbf{90}(11), 2409--2430 (2013)

\bibitem{Haentjens12}
Haentjens, T., in~'t Hout, K.J.: Alternating direction implicit finite
  difference schemes for the {H}eston--{H}ull--{W}hite partial differential
  equation.
\newblock J. Comput. Finance \textbf{16}(1), 83--110 (2012)

\bibitem{Haentjens15}
Haentjens, T., in~'t Hout, K.J.: {ADI} schemes for pricing {A}merican options
  under the {H}eston model.
\newblock Appl. Math. Finance  (2015).
\newblock \urlprefix\url{http://dx.doi.org/10.1080/1350486X.2015.1009129}

\bibitem{Heston93}
Heston, S.L.: A closed-form solution for options with stochastic volatility
  with applications to bond and currency options.
\newblock Review Financial Stud. \textbf{6}, 327--343 (1993)

\bibitem{intHout10}
in~'t Hout, K.J., Foulon, S.: A{DI} finite difference schemes for option
  pricing in the {H}eston model with correlation.
\newblock Int. J. Numer. Anal. Model. \textbf{7}(2), 303--320 (2010)

\bibitem{intHout11}
in~'t Hout, K.J., Mishra, C.: Stability of the modified {C}raig--{S}neyd scheme
  for two-dimensional convection-diffusion equations with mixed derivative
  term.
\newblock Math. Comput. Simulation \textbf{81}(11), 2540--2548 (2011)

\bibitem{intHout13}
in~'t Hout, K.J., Mishra, C.: Stability of {ADI} schemes for multidimensional
  diffusion equations with mixed derivative terms.
\newblock Appl. Numer. Math. \textbf{74}, 83--94 (2013)

\bibitem{intHout07}
in~'t Hout, K.J., Welfert, B.D.: Stability of {ADI} schemes applied to
  convection-diffusion equations with mixed derivative terms.
\newblock Appl. Numer. Math. \textbf{57}(1), 19--35 (2007)

\bibitem{intHout09b}
in~'t Hout, K.J., Welfert, B.D.: Unconditional stability of second-order {ADI}
  schemes applied to multi-dimensional diffusion equations with mixed
  derivative terms.
\newblock Appl. Numer. Math. \textbf{59}(3-4), 677--692 (2009)

\bibitem{Huang98}
Huang, J., Pang, J.S.: Option pricing and linear complementarity.
\newblock J. Comput. Finance \textbf{2}, 31--60 (1998)

\bibitem{Hull11}
Hull, J.C.: Options, futures and other derivatives.
\newblock Pearson Education, Harlow (2011)

\bibitem{Hull90}
Hull, J.C., White, A.: Pricing interest-rate-derivative securities.
\newblock Review Financial Stud. \textbf{3}, 573--592 (1990)

\bibitem{Hundsdorfer02}
Hundsdorfer, W.: Accuracy and stability of splitting with {S}tabilizing
  {C}orrections.
\newblock Appl. Numer. Math. \textbf{42}, 213--233 (2002)

\bibitem{Hundsdorfer03}
Hundsdorfer, W., Verwer, J.G.: Numerical solution of time-dependent
  advection-diffusion-reaction equations, \emph{Computational Mathematics},
  vol.~33.
\newblock Springer, Berlin (2003)

\bibitem{Ikonen04}
Ikonen, S., Toivanen, J.: Operator splitting methods for {A}merican option
  pricing.
\newblock Appl. Math. Lett. \textbf{17}(7), 809--814 (2004)

\bibitem{Ikonen07a}
Ikonen, S., Toivanen, J.: Componentwise splitting methods for pricing
  {A}merican options under stochastic volatility.
\newblock Int. J. Theor. Appl. Finance \textbf{10}(2), 331--361 (2007)

\bibitem{Ikonen07b}
Ikonen, S., Toivanen, J.: Pricing {A}merican options using {LU} decomposition.
\newblock Appl. Math. Sci. (Ruse) \textbf{1}(49-52), 2529--2551 (2007)

\bibitem{Ikonen08}
Ikonen, S., Toivanen, J.: Efficient numerical methods for pricing {A}merican
  options under stochastic volatility.
\newblock Numer. Methods Partial Differential Equations \textbf{24}(1),
  104--126 (2008)

\bibitem{Ikonen09}
Ikonen, S., Toivanen, J.: Operator splitting methods for pricing {A}merican
  options under stochastic volatility.
\newblock Numer. Math. \textbf{113}(2), 299--324 (2009)

\bibitem{Itkin11}
Itkin, A., Carr, P.: Jumps without tears: a new splitting technology for
  barrier options.
\newblock Int. J. Numer. Anal. Model. \textbf{8}(4), 667--704 (2011)

\bibitem{Jaillet90}
Jaillet, P., Lamberton, D., Lapeyre, B.: Variational inequalities and the
  pricing of {A}merican options.
\newblock Acta Appl. Math. \textbf{21}(3), 263--289 (1990)

\bibitem{Kou02}
Kou, S.G.: A jump-diffusion model for option pricing.
\newblock Management Sci. \textbf{48}(8), 1086--1101 (2002)

\bibitem{Kwon11a}
Kwon, Y., Lee, Y.: A second-order finite difference method for option pricing
  under jump-diffusion models.
\newblock SIAM J. Numer. Anal. \textbf{49}(6), 2598--2617 (2011)

\bibitem{Kwon11b}
Kwon, Y., Lee, Y.: A second-order tridiagonal method for {A}merican options
  under jump-diffusion models.
\newblock SIAM J. Sci. Comput. \textbf{33}(4), 1860--1872 (2011)

\bibitem{Lipton01}
Lipton, A.: Mathematical methods for foreign exchange.
\newblock World Scientific, Singapore (2001)

\bibitem{Marchuk90}
Marchuk, G.I.: Splitting and alternating direction methods.
\newblock Handb. Numer. Anal., I. North-Holland, Amsterdam (1990)

\bibitem{McKee70}
McKee, S., Mitchell, A.R.: Alternating direction methods for parabolic
  equations in two space dimensions with a mixed derivative.
\newblock Computer J. \textbf{13}, 81--86 (1970)

\bibitem{McKee96}
McKee, S., Wall, D.P., Wilson, S.K.: An alternating direction implicit scheme
  for parabolic equations with mixed derivative and convective terms.
\newblock J. Comput. Phys. \textbf{126}, 64--76 (1996)

\bibitem{Merton73}
Merton, R.C.: Theory of rational option pricing.
\newblock Bell J. Econom. Management Sci. \textbf{4}, 141--183 (1973)

\bibitem{Merton76}
Merton, R.C.: Option pricing when underlying stock returns are discontinuous.
\newblock J. Financial Econ. \textbf{3}, 125--144 (1976)

\bibitem{Oosterlee03}
Oosterlee, C.W.: On multigrid for linear complementarity problems with
  application to {A}merican-style options.
\newblock Electron. Trans. Numer. Anal. \textbf{15}, 165--185 (2003)

\bibitem{Peaceman55}
Peaceman, D.W., Rachford, H.H.: The numerical solution of parabolic and
  elliptic differential equations.
\newblock J. Soc. Ind. Appl. Math. \textbf{3}, 28--41 (1955)

\bibitem{Rannacher84}
Rannacher, R.: Finite element solution of diffusion problems with irregular
  data.
\newblock Numer. Math. \textbf{43}(2), 309--327 (1984)

\bibitem{Reisinger04}
Reisinger, C., Wittum, G.: On multigrid for anisotropic equations and
  variational inequalities: pricing multi-dimensional {E}uropean and {A}merican
  options.
\newblock Comput. Vis. Sci. \textbf{7}(3-4), 189--197 (2004)

\bibitem{Ruge87}
Ruge, J.W., St{\"u}ben, K.: Algebraic multigrid.
\newblock In: Multigrid methods, \emph{Frontiers Appl. Math.}, vol.~3, pp.
  73--130. SIAM, Philadelphia, PA (1987)

\bibitem{Saad86}
Saad, Y., Schultz, M.H.: G{MRES}: a generalized minimal residual algorithm for
  solving nonsymmetric linear systems.
\newblock SIAM J. Sci. Statist. Comput. \textbf{7}(3), 856--869 (1986)

\bibitem{Salmi11}
Salmi, S., Toivanen, J.: An iterative method for pricing {A}merican options
  under jump-diffusion models.
\newblock Appl. Numer. Math. \textbf{61}(7), 821--831 (2011)

\bibitem{Salmi12}
Salmi, S., Toivanen, J.: Comparison and survey of finite difference methods for
  pricing {A}merican options under finite activity jump-diffusion models.
\newblock Int. J. Comput. Math. \textbf{89}(9), 1112--1134 (2012)

\bibitem{Salmi14}
Salmi, S., Toivanen, J.: {IMEX} schemes for pricing options under
  jump-diffusion models.
\newblock Appl. Numer. Math. \textbf{84}, 33--45 (2014)

\bibitem{Seydel12}
Seydel, R.U.: Tools for computational finance.
\newblock Springer, London (2012)

\bibitem{Shreve08}
Shreve, S.E.: Stochastic calculus for finance {II}.
\newblock Springer, New York (2008)

\bibitem{Stueben01}
St{\"u}ben, K.: Algebraic multigrid: An introduction with applications.
\newblock In: Multigrid. Academic Press Inc., San Diego, CA (2001)

\bibitem{Tavella00}
Tavella, D., Randall, C.: Pricing financial instruments: The finite difference
  method.
\newblock John Wiley \& Sons (2000)

\bibitem{Toivanen08}
Toivanen, J.: Numerical valuation of {E}uropean and {A}merican options under
  {K}ou's jump-diffusion model.
\newblock SIAM J. Sci. Comput. \textbf{30}(4), 1949--1970 (2008)

\bibitem{Toivanen10a}
Toivanen, J.: A componentwise splitting method for pricing {A}merican options
  under the {B}ates model.
\newblock In: Applied and numerical partial differential equations,
  \emph{Comput. Methods Appl. Sci.}, vol.~15, pp. 213--227. Springer, New York
  (2010)

\bibitem{Toivanen10b}
Toivanen, J.: Finite difference methods for early exercise options.
\newblock In: R.~Cont (ed.) Encyclopedia of Quantitative Finance. John Wiley \&
  Sons (2010)

\bibitem{Toivanen12}
Toivanen, J., Oosterlee, C.W.: A projected algebraic multigrid method for
  linear complementarity problems.
\newblock Numer. Math. Theory Methods Appl. \textbf{5}(1), 85--98 (2012)

\bibitem{Trottenberg01}
Trottenberg, U., Oosterlee, C.W., Sch{\"u}ller, A.: Multigrid.
\newblock Academic Press Inc., San Diego, CA (2001).
\newblock With contributions by A. Brandt, P. Oswald and K. St{\"u}ben

\bibitem{Verwer99}
Verwer, J.G., Spee, E.J., Blom, J.G., Hundsdorfer, W.: A second-order
  {R}osenbrock method applied to photochemical dispersion problems.
\newblock SIAM J. Sci. Comput. \textbf{20}, 1456--1480 (1999)

\bibitem{VanderVorst92}
van~der {V}orst, H.A.: Bi-{CGSTAB}: {A} fast and smoothly converging variant of
  {B}i-{CG} for the solution of nonsymmetric linear systems.
\newblock SIAM J. Sci. Stat. Comp. \textbf{13}, 631--644 (1992)

\bibitem{Wilmott98}
Wilmott, P.: Derivatives.
\newblock John Wiley \& Sons Ltd., Chichester (1998)

\end{thebibliography}

\end{document}